\documentclass[useAMS,usenatbib]{mnras}
\usepackage{amsmath}
\usepackage{amssymb}
\usepackage{graphicx}
\graphicspath{{./figure/}}
\usepackage{color}

\newcommand{\eq}[1]{\begin{equation}\begin{split}#1\end{split}\end{equation}}

\newcommand{\eal}[1]{\begin{align}#1\end{align}}

\begin{document}
\title[Mean-Motion Resonance Capture]
{Migration of Planets Into and Out of Mean Motion Resonances in
  Protoplanetary Discs: Overstability of Capture and Nonlinear
  Eccentricity Damping}

\author[Xu, Lai \& Morbidelli]{
Wenrui Xu$^{1}$\thanks{E-mail: wenruix@princeton.edu},
Dong Lai$^{2}$ and Alessandro Morbidelli$^{3}$
\\
$^{1}$Department of Astrophysical Sciences, Princeton University, Princeton, NJ 08544, USA\\
$^{2}$Department of Astronomy, Center for Astrophysics and Planetary Science, Cornell University, 
Ithaca, NY 14853, USA\\
$^{3}$Laboratoire Lagrange, Universite Cote d’Azur, Observatoire de la Cote d’Azur, CNRS, CS 34229, 
06304 Nice, France
}

\maketitle

\begin{abstract}
A number of multiplanet systems are observed to contain planets very
close to mean motion resonances, although there is no significant
pileup of precise resonance pairs.  We present theoretical and
numerical studies on the outcome of capture into first-order mean
motion resonances (MMRs) using a parametrized planet migration model
that takes into account nonlinear eccentricity damping due to
planet-disk interaction.  This parametrization is based on numerical
hydrodynamical simulations and is more realistic than the simple
linear parametrization widely used in previous analytic studies. We
find that nonlinear eccentricity damping can significantly influence
the stability and outcome of resonance capture.  In particular,
the equilibrium eccentricity of the planet captured into MMRs become
larger, and the captured MMR state tends to be more stable compared to
the prediction based on the simple migration model.  In addition, when
the migration is sufficiently fast
or/and  the planet mass ratio is sufficiently small, 
we observe a novel phenomenon of eccentricity overshoot, where the
planet's eccentricity becomes very large before settling down to the
lower equilibrium value. This can lead to the ejection of the smaller
planet if its eccentricity approaches unity during the overshoot.
This may help explain the lack of low-mass planet companion of hot
Jupiters when compared to warm Jupiters.
\end{abstract}


\section{Introduction}

The {\it Kepler} mission has discovered thousands of exoplanets, many
of which are in multi-planet systems \citep{Batalha13,Coughlin2016}. 
The period ratio distribution of the {\it Kepler} planets shows a significant
excess of planet pairs with period ratio near mean motion resonances
(MMRs) \citep{Fabrycky14}. This excess of planets near (or in) MMRs,
together with the discovery of several resonant chain systems,
such as Kepler-223 \citep{Mills16} and TRAPPIST-1 \citep{Luger2017},
suggests that resonance capture during
disk-driven migration can be common. However, the MMR capture rate
predicted using a relatively ``clean" migration model is much higher
than the observed occurrence rate of MMRs. This discrepancy is
often explained by the disruption of MMRs by physical processes after
the resonance capture, including instability of the captured state
during disk-driven migration \citep{GoldreichSchlitchting14,DeckBatygin15,Delisle15,XuLai17}, 
tidal dissipation in planets \citep{LithwickWu12,BatyginMorbidelli13,Delisle14}, 
late time dynamical instability \citep{PuWu15,Izidoro2017}, 
and outward (divergent) migration due to planetesimal scattering
\citep{ChatterjeeFord15}.  Regardless of whether MMRs are maintained
or destroyed by any of these processes, it is important to recognize
that MMRs, even if temporarily maintained, play a significant role in
the early evolution of planetary systems and can profoundly shape
their final architectures.

A majority of the studies on the outcome of MMR capture (such as the impact of MMR on
the orbital parameters of the planets and the stability of the
resonance) include the effect of disk-driven migration using a simple parametrized
migration model, the most commonly used being that given by
\cite{GoldreichTremaine80}. The choice of this parametrized migration
model makes the equation of motion of the system relatively simple,
which is ideal for long-term numerical integrations or analytical
studies. However, this model only works well for small eccentricities 
($e\lesssim H/r$, the aspect ratio of the disk).
As we show in this paper, the eccentricities
of the planets near MMR can often lie in the regime where the
\cite{GoldreichTremaine80} result is no longer valid. 
This can impact the outcome of the resonance capture.
There are also a number of studies that includes more realistic migration
models, such as those using parametrized forcing in $N$-body
integration (e.g. \citealt{TerquemPapaloizou07,Migaszewski15}) or
using self-consistent hydrodynamics
(e.g. \citealt{Kley05,PapaloizouSzuszkiewicz05,Crida08,Zhang14,AndrePapaloizou16}). 
However, these studies tend to
focus on explaining the behaviors of particular systems and do
not survey a sufficiently large parameter space to obtain various possible outcomes.
The goal of our paper is to remedy this situation. In particular, we generalize 
previous analyses \citep{GoldreichSchlitchting14,DeckBatygin15,Delisle15,XuLai17}
by adopting a more realistic parametrization for the migration and eccentricity damping,
and examine how different model parameters affect the outcome of the MMR capture.

This paper is organized as follows. Section 2 summarizes the
parametrizations for the rates of orbit decay and eccentricity damping due to
planet-disk interactions.  In Section 3 we consider the simple case
when one of the planets is massless and study how different
parametrizations can affect the outcome of MMR capture. We find that
using the more realistic migration model can sometimes cause the
ejection of the small planet, but otherwise tend to increase the
stability of the resonance. In Section 4 we study the more realistic
case when both planets have finite masses. While most of the results
from Section 3 can be generalized, we also observe several new
phenomena that arise only when both planets have finite masses. In
addition to analytical calculations, we use 3-body integrations to
validate our results.  We conclude in Section 5 and discuss how our
results affect the architecture of multi-planet systems.


\section{Parametrizations of the rates of orbit decay and eccentricity damping}

Consider a small planet undergoing type I migration in a gaseous
disk. At low eccentricity, the rates of orbit decay and eccentricity
damping due to planet-disk interaction are approximately given by
\citep{GoldreichTremaine80}
\eal{
&\frac{\dot a}{a} = -\frac{1}{T_{m}} - \frac{2pe^2}{T_{e}},\label{eq:lin1}\\
&\frac{\dot e}{e} = -\frac{1}{T_{e}},\label{eq:lin2}
}
where $T_m,T_e$ are independent of $e$ and $T_m\sim
T_e h^{-2}$, with $h\equiv H(r)/r$ ($H$ is the disk's scale
height). The parameter $p$ characterizes the coupling between orbit decay and eccentricity damping; here we take $p=1$, which corresponds to eccentricity damping that conserves angular momentum. This is the parametrized migration model used in most studies of MMR capture 
\citep{GoldreichSchlitchting14,DeckBatygin15,Delisle15,XuLai17}.

However, this migration model is accurate only for small eccentricities,
$e\lesssim h$. For larger eccentricities, hydrodynamic
simulations \citep{Cresswell07,CresswellNelson08} show that the orbit
decay rate and eccentricity damping rate both decrease. As an empirical
fit to the numerical results, $T_m$ and $T_e$ are functions of $e/h$
given by (based on Eqs.~11 and 13 of \citealt{CresswellNelson08})
\eal{
&T_m = T_{m,0}\frac{1+(e/2.25h)^{1.2}+(e/2.84h)^6}{1-(e/2.02h)^4},\label{Tm_act}\\
&T_e = T_{e,0}\left(1-0.14\frac{e^2}{h^2}+0.06\frac{e^3}{h^3}\right),\label{Te_act}\\
&{\rm with~~}T_{m,0}\equiv \frac{t_{\rm wave}}{2.7+1.1\beta}h^{-2},~~T_{e,0}\equiv\frac{t_{\rm wave}}{0.78}.
}
Here we assume that the disk has a density profile $\Sigma(r)\propto r^{-\beta}$;
{we adopt $\beta=0$ (i.e. a disk with uniform surface density) unless otherwise specified.}
The timescale $t_{\rm wave}$ is given by (Takana \& Ward 2004)
\eq{\label{twave}
t_{\rm wave} = \frac{M_\star^2}{\Sigma a^2 m}h^4\Omega^{-1},
}
with $\Omega$ being the angular velocity of the unperturbed disk.

\begin{figure}
\center
\includegraphics[width=.8\linewidth]{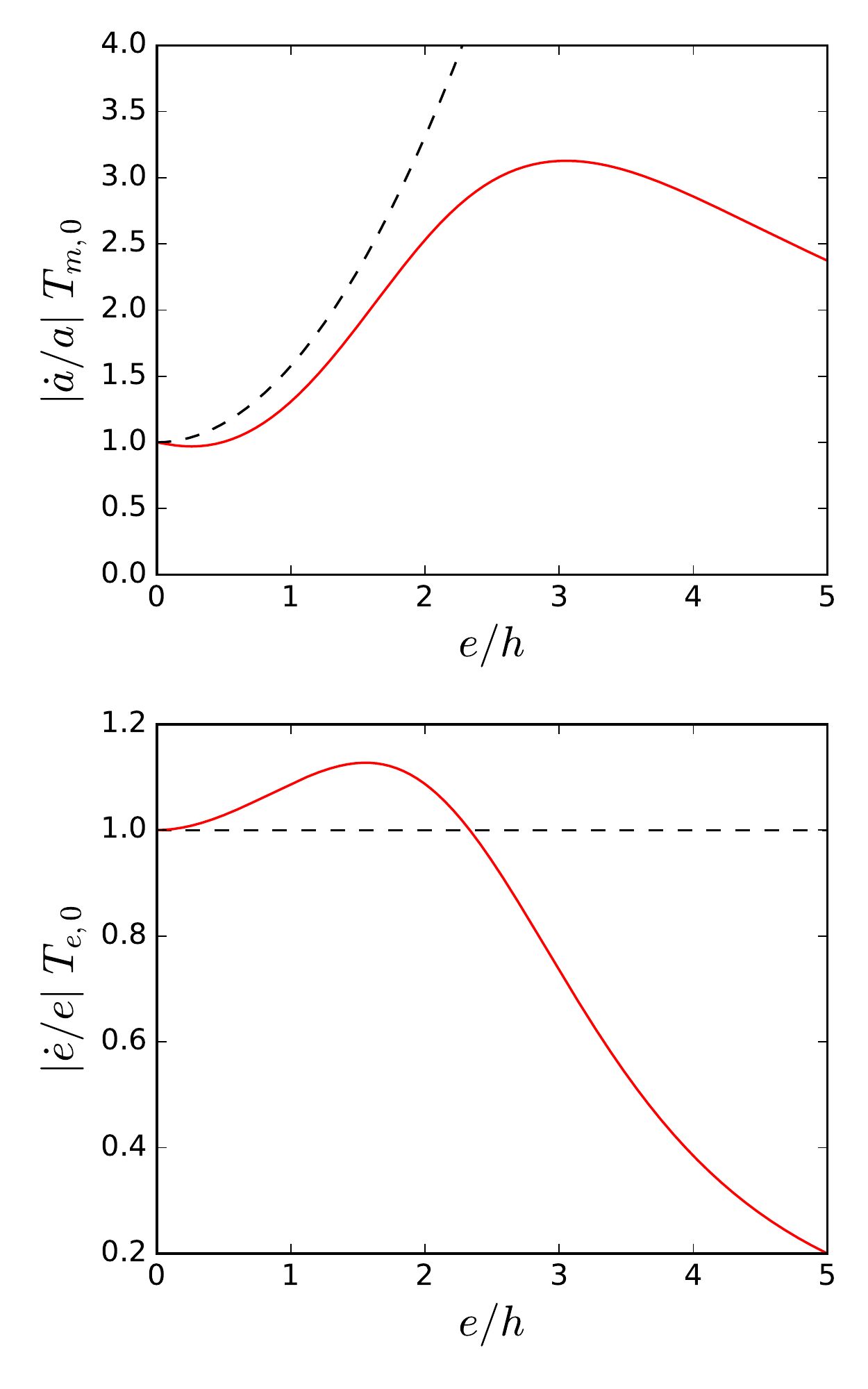}
\caption{Orbit decay and eccentricity damping rates given by the simple
  migration model (eccentricity-independent $T_e,T_m$, black dashed curves) and the realistic migration model (eccentricity-dependent $T_e,T_m$, red solid curves).
  The difference becomes prominent when $e/h\gtrsim 3$.}
\label{time}
\end{figure}

{
In this paper we compare two different migration models/
parametrizations: the ``simple'' model, with $T_m=T_{m,0}$ and
$T_e=T_{e,0}$ independent of $e$, and the ``realistic'' model, with
$T_m,T_e$ given by equations \eqref{Tm_act} and \eqref{Te_act}. The two models
are identical for $e/h\ll 1$, but can give very different orbit decay
and eccentricity damping rates when $e/h$ is large.  } This is
illustrated in Figure \ref{time}. In particular, for the realistic
migration model, the eccentricity damping rate scales as $e^{-3}$ when
$e/h\gg 1$.

\section{Outcome of MMR capture: massless inner planet}

To gain some analytical understanding to the general problem of MMR
capture with comparable mass planets, in this section we consider a
simpler case: a planet with negligible mass ($m$) perturbed by an
outer massive planet ($m'$) on a circular orbit near a first-order $j:j+1$ MMR. 
To this end, we take $T_m'$, the orbit decay timescale of the outer
planet, to be a free parameter.  This allows us to explore how the 
equilibrium eccentricity (of the inner planet), which is determined by
the net convergent migration rate, affects the outcome of the MMR
capture.  In reality, both planets undergo migration. For $m\ll m'$
and Type I migration, we expect $T_m , T_eh^{-2}\gg T_m'$.  The
results in this section should qualitatively illustrate how the
outcomes of MMR capture are affected when the realistic migration
model is applied (see Section \ref{sec:comparable_mass}).

In this section, we also assume that the planet-disk interaction is weak, so that 
$T_m,T_e,T_m'$ (note that $T_e'$ is irrelevant since the outer planet's orbit 
is always circular) are much greater than the timescale of libration, $T_{\rm res}$, given by
\eq{
T_{\rm res} \approx 0.8j^{-4/3}(\mu')^{-2/3}\frac{2\pi}{n},\label{Tres_def}
}
where $n$ is the mean motion of the inner planet and $\mu'=m'/M_\star$ 
the mass ratio between the outer planet and the star. (For the exact definition of 
$T_{\rm res}$, see Eq.~B6 in Appendix B of \citealt{XuLai17}.)
For simplicity, we assume that $T_{e,0},T_{m,0},T_m'$ remain constant (i.e. their variations 
due to the evolution of the planets' semi-major axes are ignored).

\subsection{Existence of equilibrium}

We first study the eccentricity at the equilibrium state (and whether such equilibrium state exists). Near a first-order $j:j+1$ MMR, the resonant motion conserves
\eq{
\alpha_0\equiv \alpha(1+je^2),
}
where $\alpha=a/a'<1$ is the semi-major axis ratio. When the
system undergoes convergent migration, the inner planet can be
captured into the resonance. It reaches an equilibrium state when
$d\alpha_0/dt=0$, which corresponds to
\eq{
T_{m,\rm eff}^{-1}= 2(j+1)e^2T_{e}^{-1}.\label{eq_condition}
}
Here $T_{m,\rm eff}$ (which may depend on $e$) 
is the effective convergent migration rate given
by $T_{m,\rm eff}^{-1} \equiv T_{m}'^{-1}-T_{m}^{-1}$. Note that when
the outer planet is much more massive it should migrate much faster
than the inner planet, so $T_{m,\rm eff}\approx T_{m}'$.

For the simple migration model with constant $T_e=T_{e,0}$
and $T_m=T_{m,0}$, the equilibrium always exists, with the corresponding eccentricity 
given by \citep{GoldreichSchlitchting14}
\eq{
e_{\rm eq,0} = \sqrt{\frac{T_{e,0}}{2(j+1)T_{m,\rm eff, 0}}} \simeq \sqrt{\frac{T_{e,0}}{2(j+1)T_{m}'}},
}
where $T_{m,\rm eff,0}^{-1} \equiv T_{m}'^{-1}-T_{m,0}^{-1}$.

\begin{figure*}
\center
\includegraphics[width=.9\linewidth]{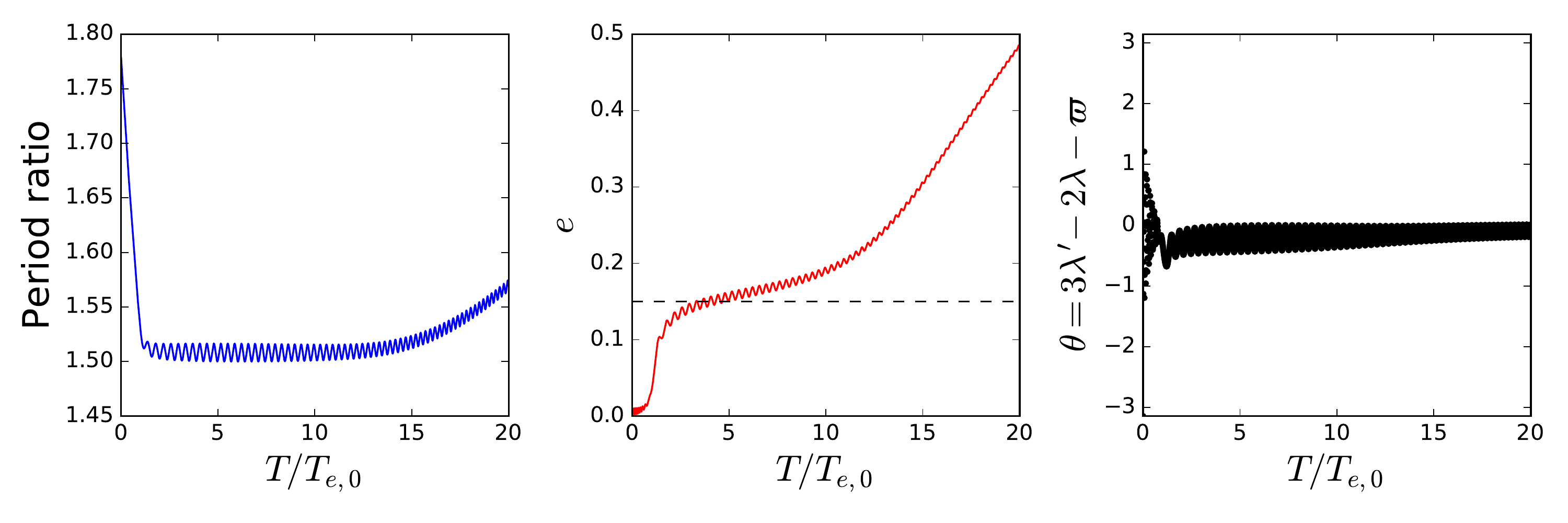}
\caption{Evolution of a system near 2:3 MMR with $h=0.05$, 
$T_{m}' = 10  T_{e,0}$, and the 
outer planet mass $\mu'=m'/M_\star = 10^{-3}$.
Left panel: period ratio. Center panel: eccentricity of
  the inner planet. The black dashed line shows $e=3h$ for
  reference. Right panel: resonant angle. The equilibrium does not
  exist and the planet's eccentricity grows unboundedly as the system
  goes deeper into the resonance. This increasing
  eccentricity should ultimately cause ejection or collision of the planet.}
\label{T10}
\end{figure*}

However, for the realistic migration model with eccentricity-dependent $T_e$ and $T_m$, the right-hand side
of \eqref{eq_condition} has a finite maximum value because for
$e\gtrsim$ a few $h$, $e^2T_{e}^{-1}\propto e^{-1}$ decreases as $e$
increases. Therefore, the equilibrium may not exist when the outer
planet's migration is too fast. The maximum value of
$e^2T_{e}^{-1}$ occurs at $e\simeq 3h$, thus the equilibrium 
ceases to exist when
\eq{
T_{m}'^{-1} \gtrsim 2(j+1) T_{e,0}^{-1}(3h)^2.\label{Tm_cond}
}
Figure \ref{T10} gives an example of the evolution of the system when the
equilibrium of resonance capture does not exist.

\subsection{Stability of capture}

The migration model can also affect the stability of the captured (equilibrium) state.

For the simple migration model, the stability of the equilibrium state
has been studied by \cite{GoldreichSchlitchting14}.  Under the
assumption that planet-disk interaction is weak, the behavior of the
system depends only on the ratio $\mu'/ e_{\rm eq,0}^3$, where $\mu'=m'/M_\star$
and $e_{\rm eq,0}$ is the previously defined equilibrium eccentricity. 
The equilibrium is stable when the outer planet is sufficiently massive
(with $\mu'\gtrsim e_{\rm eq,0}^3$); in this case the resonant angle
librates with small amplitude.  
For $\mu'\sim e_{\rm eq,0}^3$, the libration amplitude
saturates at a finite value, and the system stays in resonance.
For a less massive outer planet (with $\mu'\lesssim e_{\rm eq,0}^3$), the
equilibrium state is overstable (i.e. the amplitude of libration
increases with time) and the system eventually escapes from resonance.

For the realistic migration model, however, the stability of the
equilibrium state depends on not only $\mu'/e_{eq,0}^3$ but also
$e_{\rm eq,0}/h$; the latter parameter characterizes how significantly
the system is affected by including the eccentricity dependence in the
migration model.

\begin{figure}
\centering
\includegraphics[width=\linewidth]{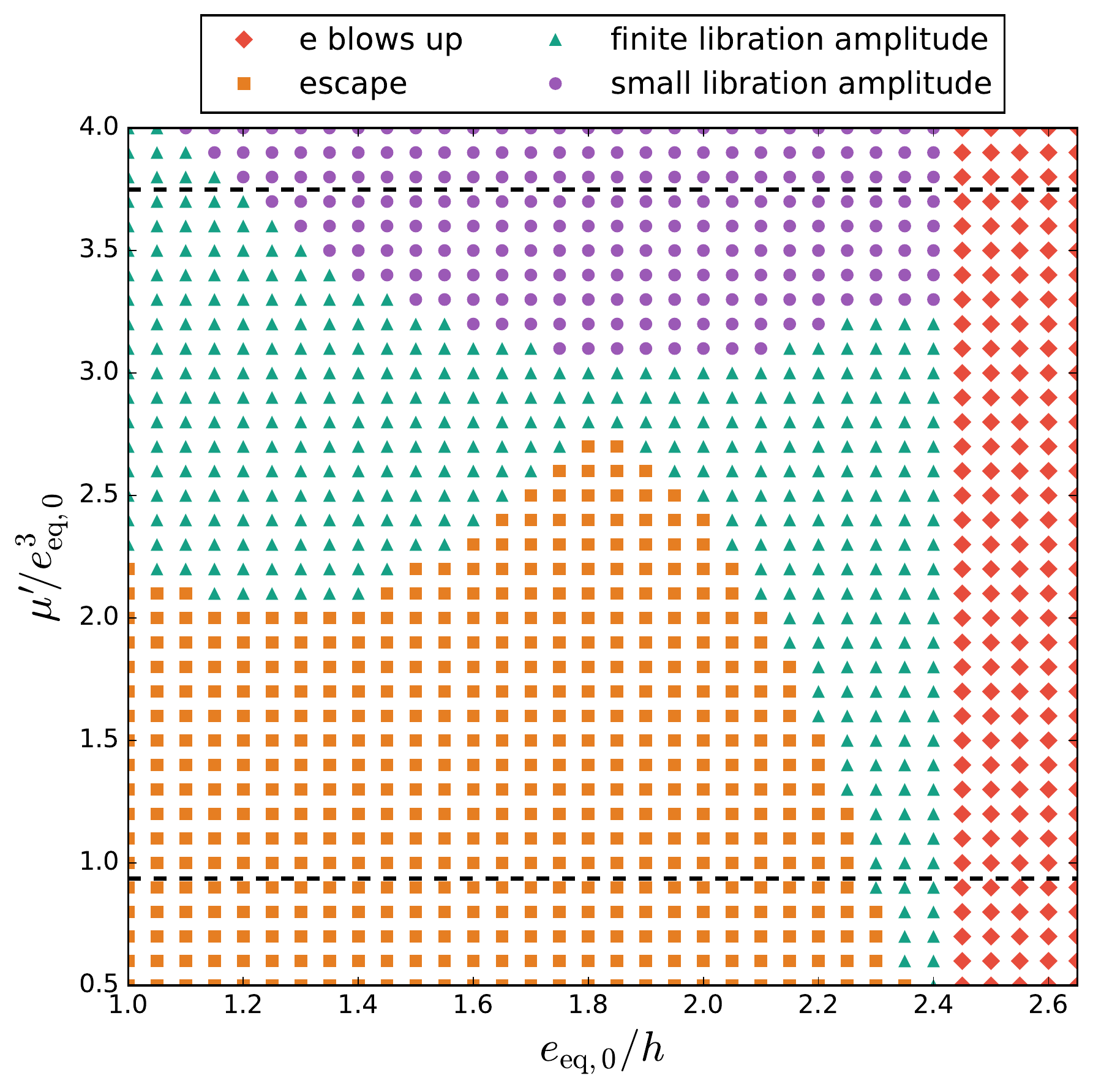}
\caption{Regimes of different behaviors in the 
$e_{eq,0}/h$ -  $\mu'/e_{eq,0}^3$ parameter space for a 2:3 MMR. The two dashed
  lines mark the analytical estimates for the boundary between
  libration with small amplitude, libration with finite amplitude and
  escape given by \citet{GoldreichSchlitchting14} for the simple migration model.}
\label{paramSp}
\end{figure}

Figure \ref{paramSp} plots the regimes of different behaviors in the
$e_{eq,0}/h$ - $\mu'/e_{eq,0}^3$ parameter space for a 2:3 MMR when
the realistic migration model is applied.  We integrate the equation
of motion derived from the resonance Hamiltonian
(see, e.g., Appendix B of Xu \& Lai 2017), and include the
dissipative terms associated with migration and eccentricity
damping.\footnote{ Direct integration of the equation of motion is
  necessary because the outcome when the equilibrium state is
  overstable (whether the libration saturates at a finite amplitude,
  or the system eventually escapes the resonance) cannot be obtained
  from linear stability analysis of the equilibrium state.}
We find that there are {\it four possible outcomes/behaviors:}

(i) When $e_{eq,0}$ is larger than $2.4h$,
the equilibrium state of resonance capture does not exist because the
eccentricity damping is too weak to balance the eccentricity excitation due to
resonant interaction, and the planet's eccentricity $e$
grows unboundedly until the system becomes unstable (red diamonds
in Fig.~\ref{paramSp}). 

(ii)-(iv) When $e_{eq,0}$ is small enough to allow the existence of an
equilibrium state, this equilibrium can be stable or
overstable. When it is stable, the system exhibits small libration around
the equilibrium state with the libration amplitude converging to zero
(purple circles in Fig.~\ref{paramSp}).
When it is overstable, the system can either end up
in a stable state with a finite libration amplitude (green triangles) or
exit the resonance with damped eccentricity (orange squares). 
Only these three behaviors are possible in the simple migration model.

Although Figure \ref{paramSp} refers to the 2:3 MMR, we find that 
the results for other first-order MMRs are qualitatively similar.

Three-body simulations (see below) show that the results obtained from
the resonant Hamiltonian in Figure \ref{paramSp} are qualitatively
correct, with tolerable error for the boundaries between different
behaviors. Note that the boundaries between the last three behaviors
(stable libration with finite and small amplitude, and escape) depend
sensitively on the migration model, since the stability of the
equilibrium is affected by the derivatives of $T_e$ and
$T_m$.\footnote{ {One can see this by considering how the
    stability of the equilibrium point is calculated. The stability is
    determined by the eigenvalues of a matrix with entries of the form
    $\partial_x(dy/dt)$, where $x,y$ can be either $a$ or $e$. These entries
    depend not only on the values of $T_e$ and $T_m$ but also on their derivatives 
with respect to $a$ or $e$.}}

Figure \ref{paramSp} differs from the result based on the simple
migration model (e.g. \citealt{GoldreichSchlitchting14}) in several
aspects. First, as noted above, there exists a new regime where the
planet's eccentricity can grow unboundedly because of the decrease of
eccentricity damping rate for $e\gtrsim h$. Second, near the boundary
of this ``eccentricity blowing up'' regime ($2.3\lesssim
e_{eq,0}/h\lesssim 2.45$), the stable finite-amplitude libration
regime occupies a large parameter space; in particular, the system can
stay in resonance with a finite-amplitude libration even when
$\mu'/e_{eq,0}^3$ is as small as $0.6$ (by contrast, the simple
migration model would predict the system escape from the resonance due
to overstability).  Third, the boundaries between the different
regimes, even at $e_{eq,0}/h\lesssim 2$ (for which $T_m$ and $T_e$
deviate little from the simple model), are significantly distorted due
to the use of the more realistic migration model, showing that these
boundaries are indeed sensitive to the migration model (and disk
parameters).  Note that for low eccentricity ($e_{\rm eq,0}/h\lesssim
1$) our result may not be accurate given that the fitting used to
obtain equations \eqref{Tm_act} and \eqref{Te_act} may introduce nontrivial
error in the derivatives of $T_m$ and $T_e$ when $e\to 0$.  Therefore,
we do not expect our model to recover the analytical result of
\cite{GoldreichSchlitchting14} for small $e_{\rm eq}$, and result for
$e_{\rm eq}/h<1$ is not shown in Figure \ref{paramSp}.

\begin{figure}
\centering
\makebox[\linewidth][c]{\includegraphics[width=1.1\linewidth]{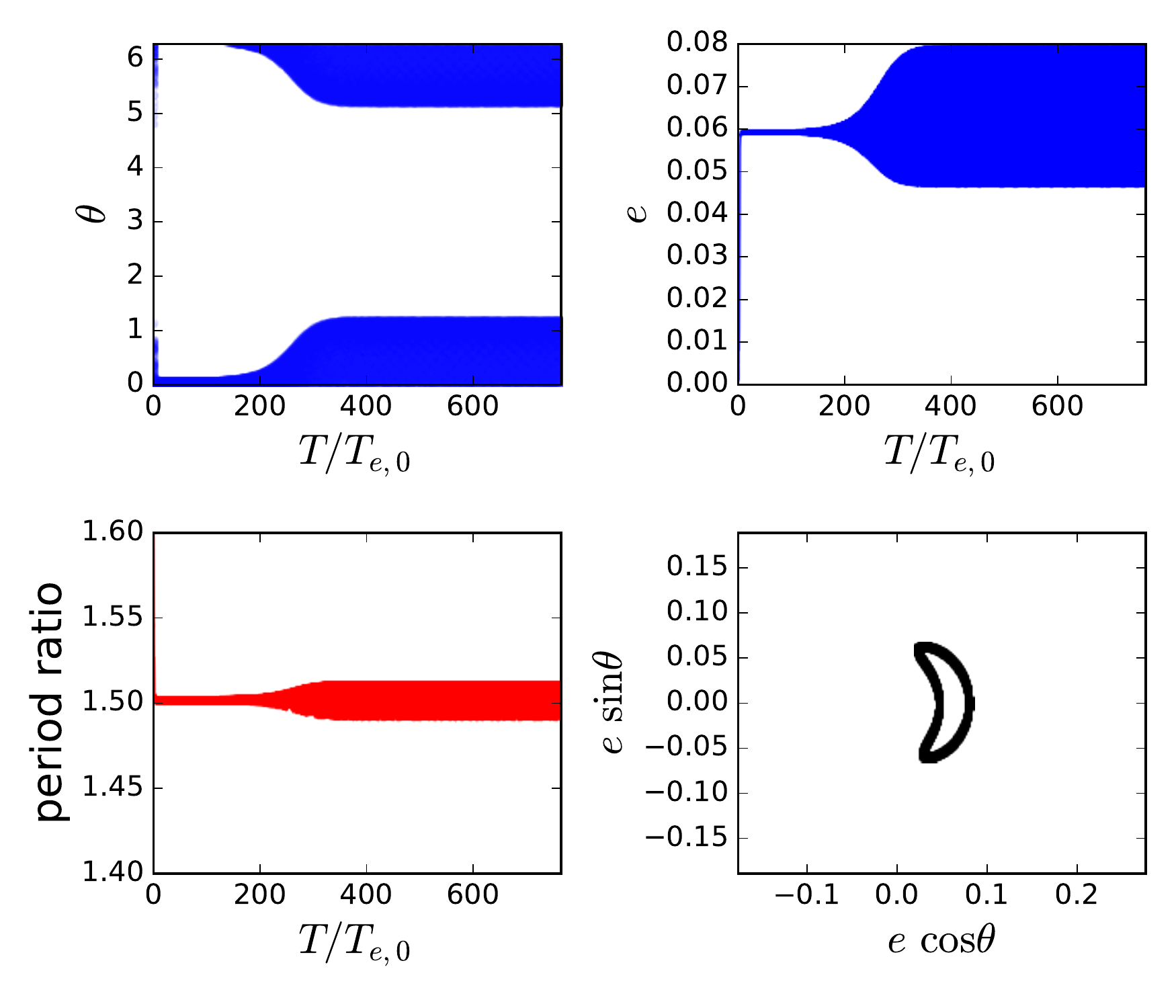}}
\caption{Evolution of a system captured into the 2:3 MMR with
  $e_{{\rm eq},0}/h = 2.3$ and $\mu'/e_{{\rm eq},0}^3 = 1$. The four panels plot
  the inner planet's resonant angle, eccentricity, the two planet's
  period ratio, and the inner planet's trajectory in $e\cos\theta$ -
  $e\sin\theta$ phase space after reaching the equilibrium. 
  The system ends up in a stable state with finite libration
  amplitude.}
\label{example1}
\end{figure}

\begin{figure}
\centering
\makebox[\linewidth][c]{\includegraphics[width=1.1\linewidth]{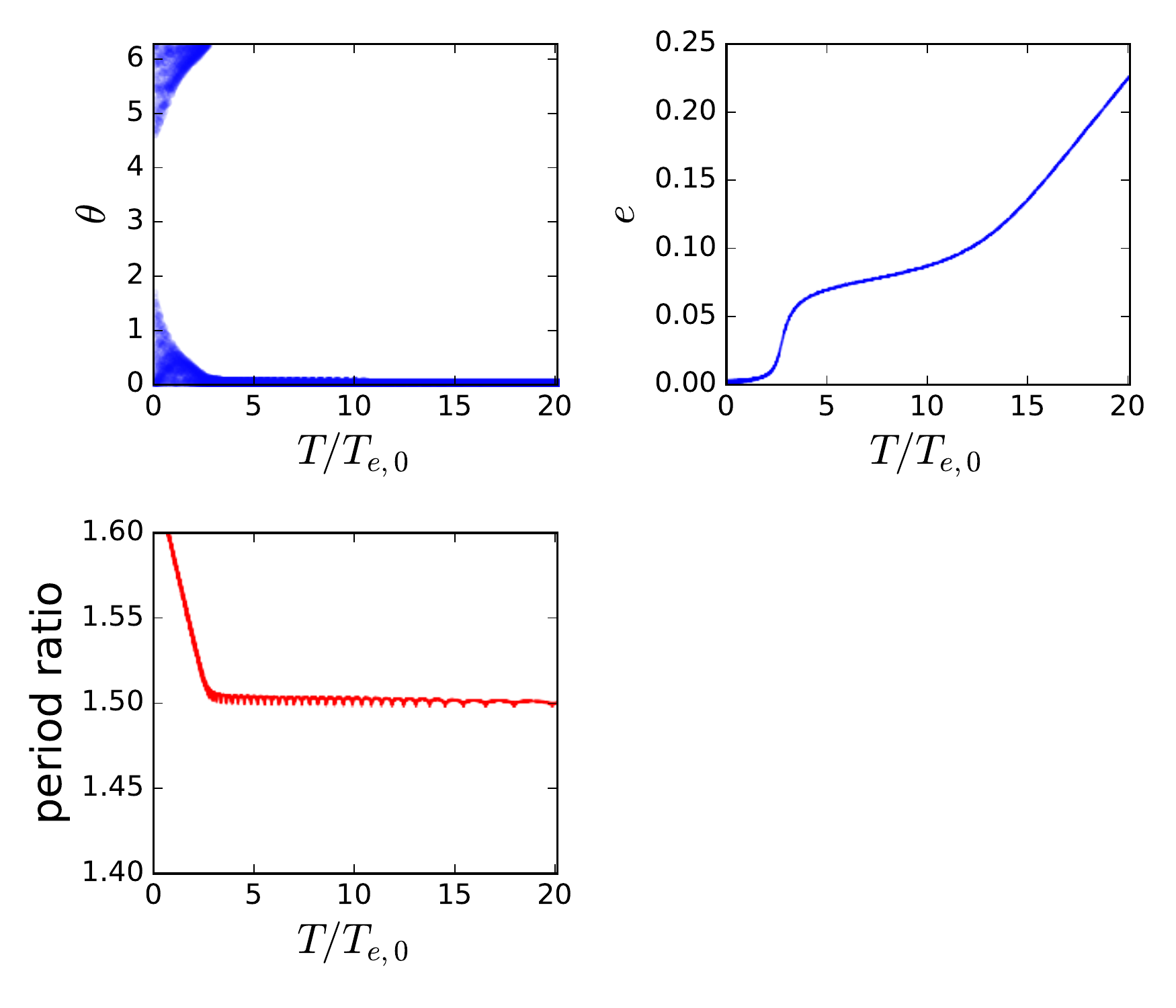}}
\caption{Same as Fig.~\ref{example1}, but with $e_{{\rm eq},0}/h = 2.5$
  and $\mu'/e_{{\rm eq},0}^3 = 1$. Equilibrium no longer exists and the
  planet's eccentricity grows unboundedly. The system eventually
  becomes unstable and the inner planet gets ejected shortly after $e$
  reaches $\sim 0.8$.}
\label{example2}
\end{figure}

\begin{figure}
\centering
\makebox[\linewidth][c]{\includegraphics[width=1.1\linewidth]{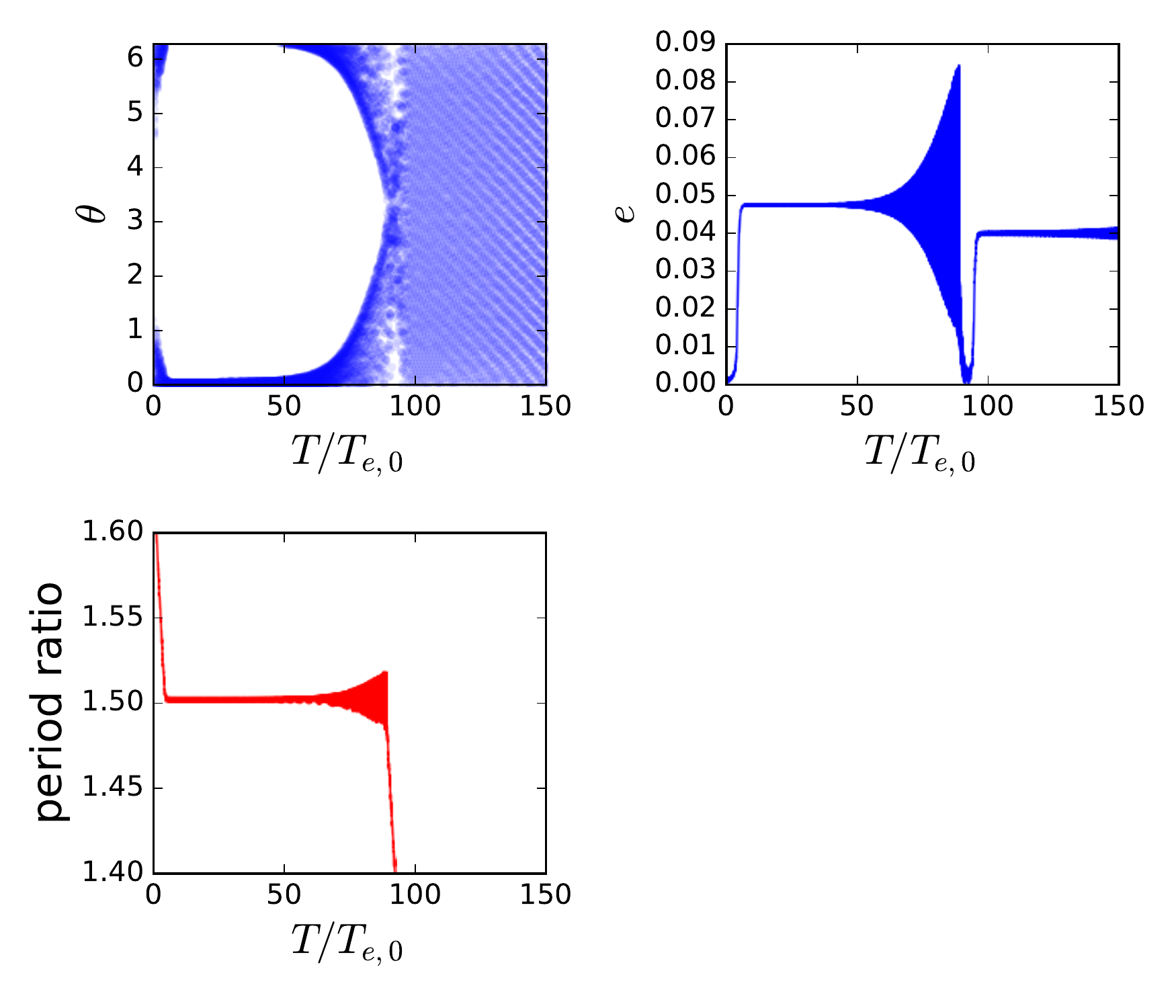}}
\caption{Same as Fig.~\ref{example1}, but with $e_{{\rm eq},0}/h = 2$ and
  $\mu'/e_{{\rm eq},0}^3 = 1$. The planet escapes from the resonance due to
  the overstability of the equilibrium state. Note that soon after the
  planet exits the 2:3 MMR, it gets captured into a 3:4 MMR.}
\label{example3}
\end{figure}

\begin{figure}
\centering
\makebox[\linewidth][c]{\includegraphics[width=1.1\linewidth]{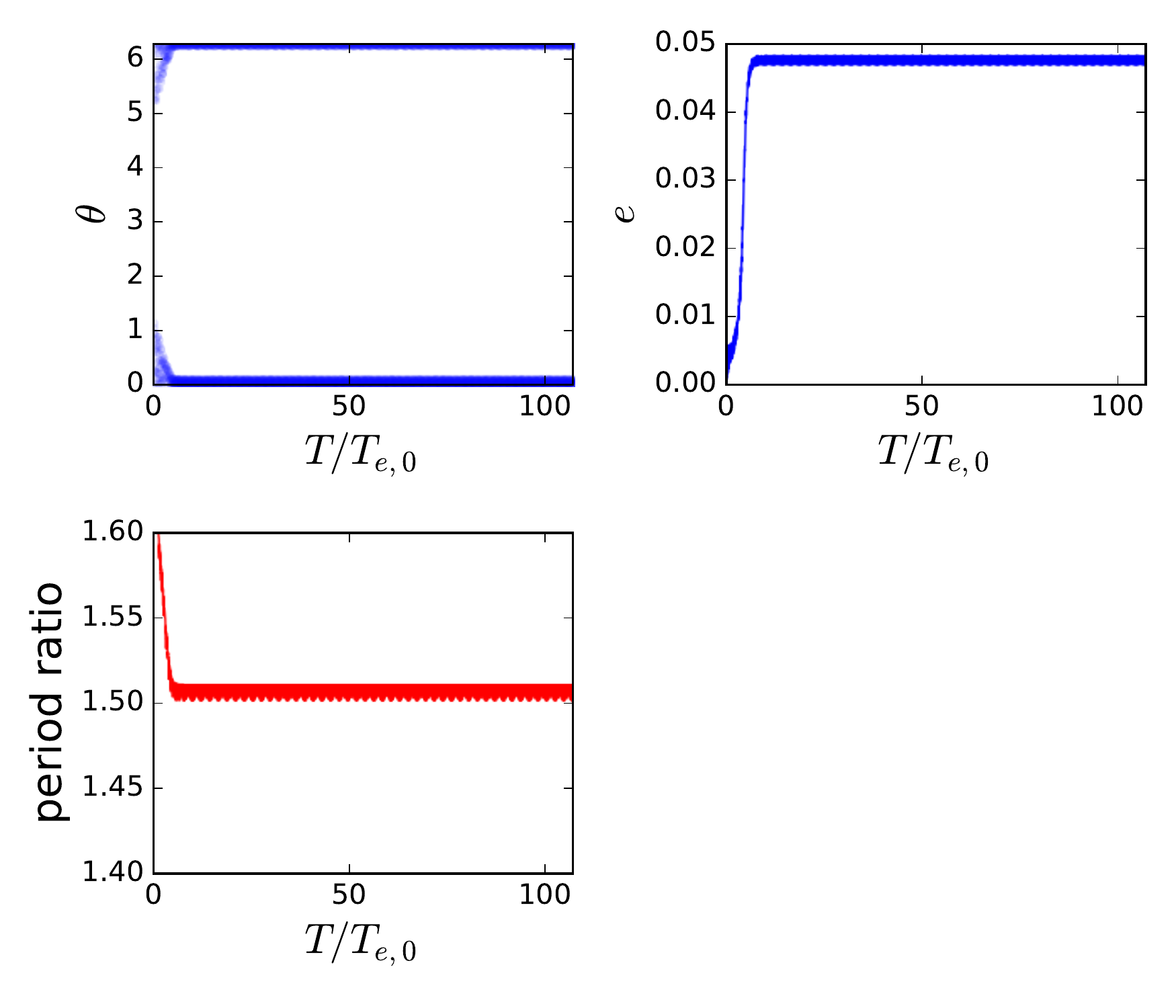}}
\caption{Same as Fig.~\ref{example1}, but with $e_{{\rm eq},0}/h = 2$ and
  $\mu'/e_{{\rm eq},0}^3 = 3.5$. The equilibrium state is stable and the
  libration of the resonant angle stays small.}
\label{example4}
\end{figure}

Figures \ref{example1}-\ref{example4} show the behavior of the system
in each regime depicted in Fig.~\ref{paramSp}.
These results are obtained by doing 3-body integrations
using the \textit{MERCURY} code \citep{Chambers99}, with $h=0.025$,
$a'=1$~au and 
$T_{e,0}=10T_{\rm res}$ 
[with $T_{\rm res}$ given by equation \eqref{Tres_def}].
The other parameters of the system
can be solved to match the given $e_{eq,0}/h$ and $\mu'/e_{eq,0}^3$ values. 
In practice, to avoid having the planets migrate too far inward 
during the integration (which will make it necessary to choose a much smaller timestep 
to account for the planet's short orbital period),
we fix the outer planet and let the inner planet's semi-major
axis increase at the rate $\dot a/a=
-1/T_{m}-2e^2/T_{e}+1/T_{m}'$ 
--- Note that this parameterized treatment is necessary because the overstability 
timescale of the equilibrium state can be $\gtrsim 10 T_{m}'$ in many cases.

\section{Outcome of MMR capture: two massive planets}\label{sec:comparable_mass}

To apply our results to realistic systems, it is important to study
the case where both planets have finite masses.  As we will show in
this section, the perturbation on the more massive planet from the
smaller planet can qualitatively affect the outcome of resonance
capture even when the mass ratio is very small. We will also discuss
the effect of strong eccentricity damping rate and non-adiabatic
evolution due to fast migration.

\subsection{Existence and location of equilibrium}

Consider two planets near a $j:j+1$ MMR, with both planets having
finite masses. Let the inner (outer) planet have mass $m_1$ ($m_2$)
and semi-major axis $a_1$ ($a_2$).\footnote{The notation is different
  from Section 3 in order to emphasize the fact that both
  planets have finite masses.}  The Hamiltonian of the system, to
first order in eccentricity and with all non-resonant terms averaged out,
is given by
\eq{
H = &-\frac{GM_\star m_1}{2a_1}-\frac{GM_\star m_2}{2a_2}\\
&-\frac{Gm_1m_2}{a_2}\left(f_{j+1,27}\,e_1\cos\theta_1+\tilde f_{j+1,31}\,e_2\cos\theta_2\right).\label{eq:Ham}
}
Here $\alpha=a_1/a_2$, and $f_{m,n}$ are functions of $\alpha$ (evaluated at $\alpha_0\equiv [j/(j+1)]^{2/3}$) given in Appendix B of \citet{MurrayDermott99}, with $\tilde f_{j+1,31}\equiv f_{j+1,31}-\delta_{j,1}2\alpha_0$.
The interaction between the two planets conserves the total angular momentum
\eq{
\mathcal L\equiv \Lambda_1\sqrt{1-e_1^2}+\Lambda_2\sqrt{1-e_2^2},
}
where $\Lambda_i=m_i\sqrt{GM_\star a_i}$.
The Hamiltonian \eqref{eq:Ham} also admits a second constant of motion \citep{MichtchenkoFerraz-Mello2001},
\eq{
\mathcal K\equiv \frac{j+1}{j}\Lambda_1 + \Lambda_2.
}
Combining the two ($\mathcal L$ and $\mathcal K$) produces a conserved quantity $\eta$, given by
\eq{
\eta&\equiv -2(q\alpha_0^{-1}+1)\left[\frac{\mathcal L}{\mathcal K}-\left(\frac{\mathcal L}{\mathcal K}\right)_{\alpha=\alpha_0,e_i=0}\right]\\
&\approx\frac{q}{j\alpha_0^{1/2}(q\alpha_0^{-1}+1)}(\alpha-\alpha_0) + \alpha_0^{1/2}q e_1^2 + e_2^2,\label{eq:eta_def}
}
where $q=m_1/m_2$ is the mass ratio, and $\alpha_0 = [j/(j+1)]^{2/3}$
is the semi-major axis ratio at resonance.  In the second line of
equation \eqref{eq:eta_def} we have expanded the result to the lowest order in
$(\alpha-\alpha_0)$ and $e_i^2$.  
The parameter $\eta$ characterizes how deep the system is inside the resonance when captured:
For larger $\eta$, the system is deeper inside the resonance, 
and the fixed point (libration center) of the system corresponds to larger eccentricities.

Consider the evolution of $e_i,\varpi_1-\varpi_2$ and $\eta$. At the
equilibrium state, $\varpi_1-\varpi_2$ is constant because the
resonant angles $\theta_i\equiv (j+1)\lambda_2-j\lambda_1-\varpi_i$
are constant; $\eta$, which is a function of $\alpha$ and
$e_i$, should also be constant because $\alpha,e_i$ are constant.
Therefore, $e_i$ and $\theta_i$ at the equilibrium state can be solved from the following 
equations:\footnote{
Another method is to directly solve for the equilibrium state by linking the evolution of all quantities to that of $R\equiv a_2/a_1$, and imposing that $\dot R=0$, all the while considering the torques exerted by the disk on the planets which result from the migration model (Pichierri et al. 2018, in preparation).
Our approach makes it easier to analyse how the equilibrium eccentricities are 
affected by using different migration models.}
\eal{
&\frac{de_1}{dt} = -\mu_2n_1\alpha_0 f_{j+1,27}\sin\theta_1-\frac{e_1}{T_{e,1}}=0,\label{eq:e1}\\
&\frac{de_2}{dt} = -\mu_1n_2\tilde f_{j+1,31}\sin\theta_2-\frac{e_2}{T_{e,2}}=0,\label{eq:e2}\\
&\frac{d(\varpi_1-\varpi_2)}{dt} \propto f_{j+1,27}e_2\cos\theta_1-\tilde f_{j+1,37}e_1q\alpha_0^{1/2}\cos\theta_2 =0,\label{eq:varpi}\\
&\frac{d\eta}{dt} =\frac{q\alpha_0^{1/2}}{j(q\alpha_0^{-1}+1)}\left(\frac{1}{T_{m,2}}-\frac{1}{T_{m,1}}+\frac{2e_1^2}{T_{e,1}}-\frac{2e_2^2}{T_{e,2}}\right)\nonumber\\
&~~~~~~~~-q\alpha_0^{1/2}\frac{2e_1^2}{T_{e,1}}-\frac{2e_2^2}{T_{e,2}} =0.\label{eq:eta}
}
Note that since $\eta$ is conserved in the absence of dissipation (planet-disk interaction),
equation (\ref{eq:eta}) only includes contributions from planet-disk interactions.
Equation \eqref{eq:eta} can be interpreted physically as that
convergent migration tends to push the system deeper into resonance
(i.e. increases $\eta$ and eccentricities) and while eccentricity damping 
(from planet-disk interaction) counters the effect
of migration. Equilibrium is reached ($\eta$ ceases to evolve) when
migration and eccentricity damping balance each other.

\subsubsection{Weak eccentricity damping}

First consider the case when the eccentricity damping is weak,
i.e. $\mu_2n_1\gg e_1/T_{e,1}$ and $\mu_1n_2\gg e_2/T_{e,2}$. In this
case, $|\cos \theta_i| \approx 1$, and equation \eqref{eq:varpi} gives
$e_1/e_2$ at the equilibrium. Note that 
$e_1/e_2\sim q^{-1}=m_2/m_1$ 
and is independent of $T_{e,i}$ and $T_{m,i}$. With $e_1/e_2$ known,
and with $T_{e,i}, T_{m,i}$ as a function of $e_i$ (see Section 2), we
can solve equation \eqref{eq:eta} to obtain $e_i$ at the equilibrium.

\begin{figure*}
\centering
\includegraphics[width=\linewidth]{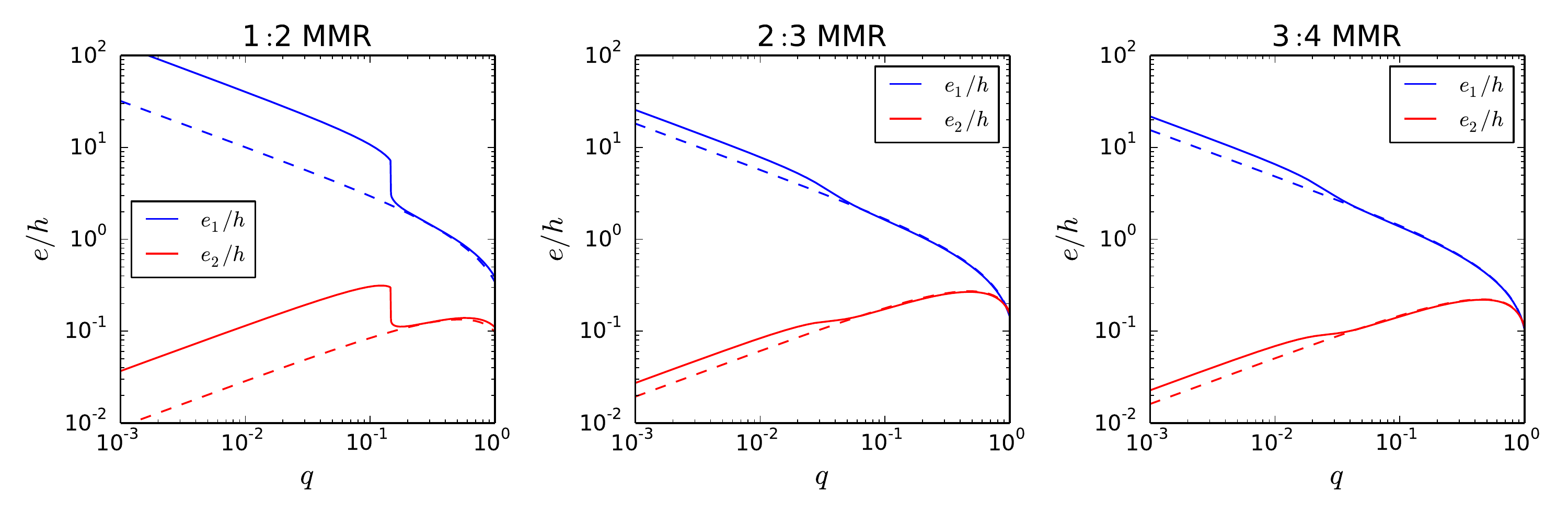}
\caption{Equilibrium eccentricities vs mass ratio $q=m_1/m_2$ of the two planets captured into a
  first-order MMR in a disk with uniform surface density ($\beta=0$ in
  $\Sigma\propto r^{-\beta}$).  Note that the $e_i/h$ value depends only
  on $q$ and is independent of the total mass of the planets and
  $h$. Only the $q<1$ region is plotted since convergent migration requires
  $q\lesssim 1$. The blue (red) curve shows the eccentricity of the inner
  (outer) planet. The solid (dashed) curve shows the eccentricity for the
  realistic (simple) migration model. These results are calculated
  under the assumption that eccentricity damping is weak.}
\label{EqlibEcc_analytic}
\end{figure*}

Figure \ref{EqlibEcc_analytic} shows the equilibrium eccentricities of
the two planets calculated using the above method.  For the simple
migration model, equations \eqref{eq:varpi} and \eqref{eq:eta} give $e_{1,\rm eq}
\sim q^{-1/2}h$ and $e_{2,\rm eq}\sim q^{1/2}h$. At the equilibrium,
the $e_1^2$ terms and the $e_2^2$ terms in equation \eqref{eq:eta} are comparable
when $q\lesssim 1$: The eccentricity terms in the first line of \eqref{eq:eta} are comparable to or smaller than the corresponding eccentricity terms in the second line when $q\lesssim 1$, and $q e_1^2 / T_{e,1} \sim e_2^2 / T_{e,1}$ given that $T_{e,1}/T_{e,2}\sim q^{-1}$.

For the realistic migration model, the result is similar to that of
the simple migration model when $q$ is relatively large ($q\gtrsim
0.15, 0.04$ and $0.03$ for the 1:2, 2:3 and 3:4 MMR respectively).  When
$q$ is smaller, however, $e_{1,\rm eq}$ exceeds $\sim 3h$ and
the damping rate $T_{e,1}^{-1}$ is reduced.
Therefore, the equilibrium eccentricities of the planets must increase in order to
satisfy equation \eqref{eq:eta}.

The equilibrium always exists when both planets have finite masses,
although it may correspond to $e_{1,\rm eq}\gtrsim 1$,
which implies that the smaller planet can be ejected due to
instability before reaching the equilibrium. This is very different
from the ``massless inner planet'' case considered in Section 3, 
where the equilibrium state may not exist. The
reason of such a difference is that while the eccentricity of the
smaller planet can exceed $3h$, the eccentricity of the more massive planet
always remains well below $3h$, so that the eccentricity damping from the
more massive planet is able to balance migration, ensuring the
existence of an equilibrium state.

\begin{figure}
\centering
\includegraphics[width=\linewidth]{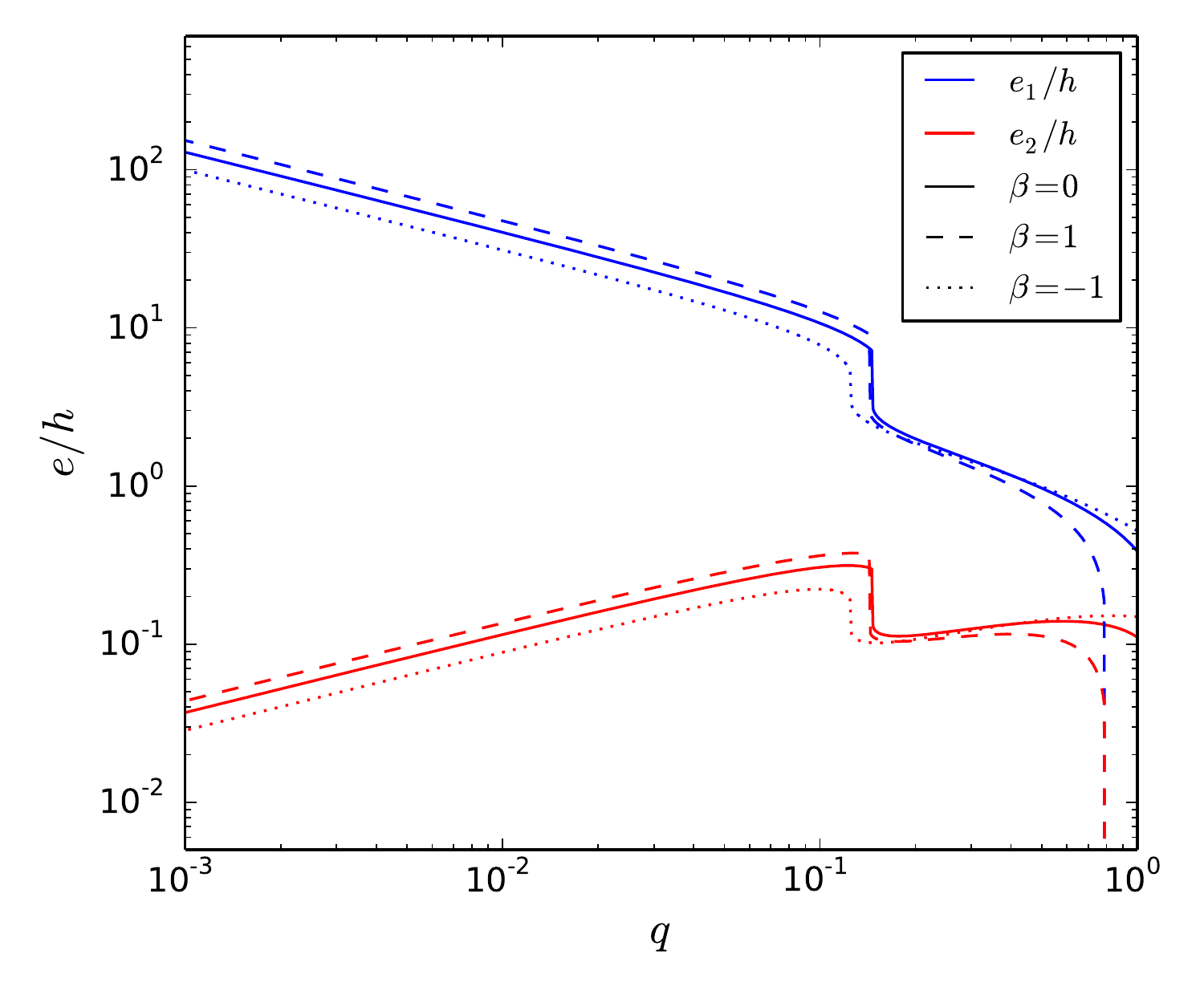}
\caption{Equilibrium eccentricities of the two planets captured into a
  1:2 MMR in disks with different $\beta$. The blue (red) curves show
  the eccentricities of the inner (outer) planet. The solid, dashed and
  dotted curves are the eccentricities for $\beta=0,1$ and $-1$
  respectively. The results are similar for different values of $\beta$.}
\label{EqlibEcc_beta}
\end{figure}

The equilibrium eccentricities also depend on the density profile of
the disk, which is characterized by the parameter $\beta$ [assuming
that the disk has $\Sigma(r)\propto r^{-\beta}$; note that we adopt
$\beta=0$ everywhere else in this paper].  Figure
\ref{EqlibEcc_beta} shows that the equilibrium eccentricities of the
planets depend weakly on $\beta$.

\subsubsection{Effect of strong eccentricity damping}\label{subsubsec:finite_damping}

\begin{figure}
\centering
\includegraphics[width=\linewidth]{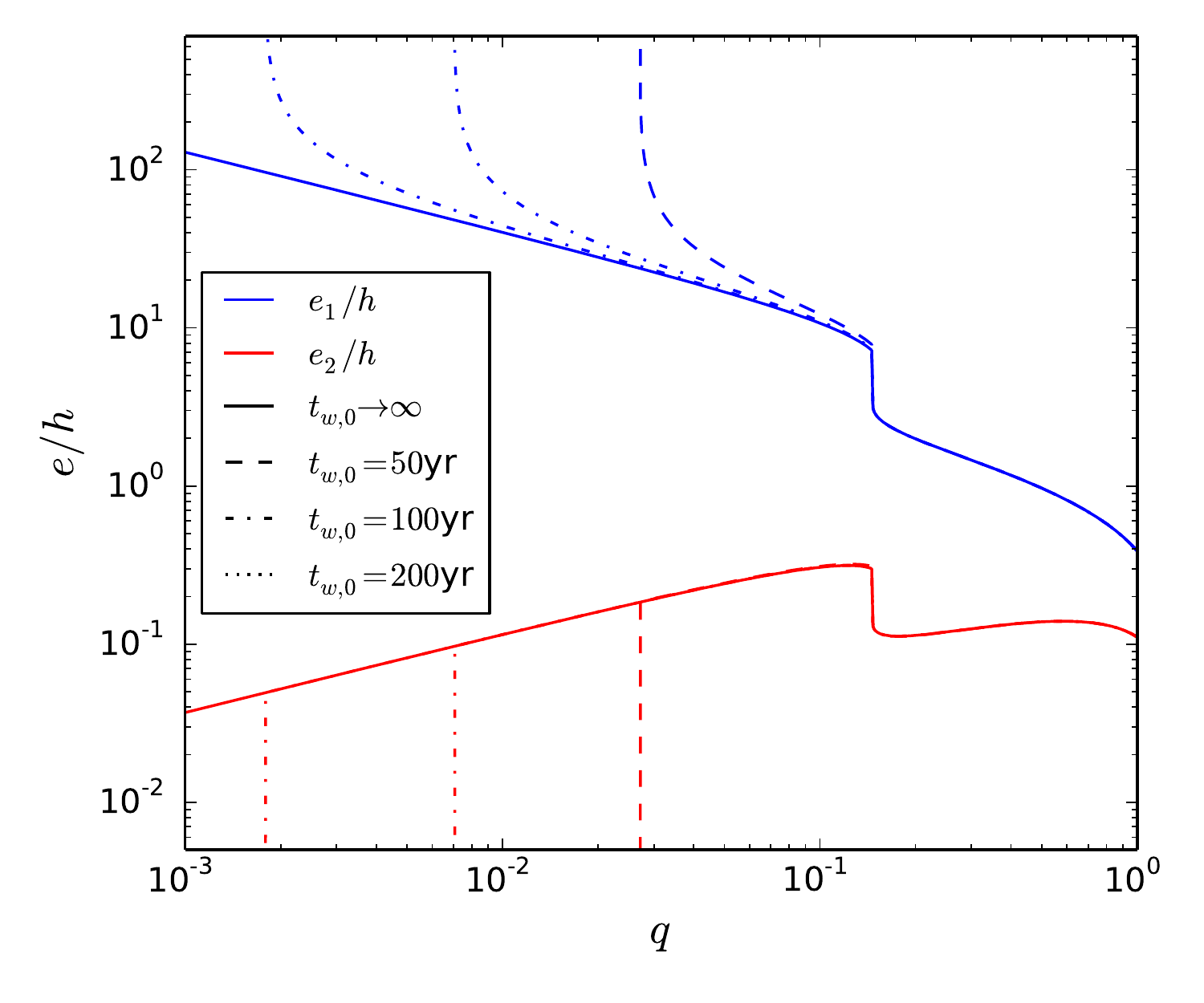}
\caption{Equilibrium eccentricities of the two planets with
  $\mu_1+\mu_2=10^{-3}, M_\star=1M_\odot$ and $a_2=1$~au captured into
  the 1:2 MMR in a disk with uniform surface density, for different
  strengths of eccentricity damping. The eccentricity damping rate is
  characterized by $t_{w,0}$, defined as $t_{\rm wave}$ (see
  Eqs.~\ref{Tm_act}-\ref{twave}) evaluated at $m=m_1+m_2$ and
  $a=a_2$. The blue (red) curves show the eccentricity of the inner
  (outer) planet. Different line styles correspond to different
  $t_{w,0}$, with the solid curves corresponding to very slow
  migration ($t_{w,0}\to \infty$).  }
\label{EqlibEcc_tw}
\end{figure}

When $q$ is small, the resonant perturbation from the inner planet is no longer much stronger than
the eccentricity damping of the outer planet, and the second term in \eqref{eq:e2} can no longer
be ignored.
In this regime, the equilibrium eccentricities can be
significantly affected when the realistic migration model is applied.

For sufficiently small $q$ (which gives large $e_1/h$), the terms proportional to $e_1^2$ 
in equation \eqref{eq:eta} are negligible, so $e_{2,\rm eq}\sim q^{1/2} h$ can be determined
directly from \eqref{eq:eta} and is independent of the strength of
eccentricity damping. 
Meanwhile, equations \eqref{eq:e2} and \eqref{eq:varpi}
suggest that for smaller $\mu_1$ (or larger $1/T_{e,2}$),
$|\sin\theta_2|$ increases, $|\cos\theta_2|$ decreases, and $e_{1,\rm eq}/e_{2,\rm eq}$ 
increases.  In particular, when $\mu_1$ is sufficiently small (i.e. $\mu_1n_2\sim e_{2,\rm eq}/T_{e,2}\sim q^{1/2}h/T_{e,2}$), 
$|\cos\theta_2|\to 0$ and $e_{1,\rm eq}/e_{2,\rm eq}$ diverges.
Since $e_{2,\rm eq}$ is finite, this means that $e_{1,\rm eq}$
diverges (i.e. ejection or collision of the smaller planet should
happen before the equilibrium is reached.)

Figure \ref{EqlibEcc_tw} (based on numerical calculations of the
equilibrium eccentricities) demonstrates this effect.  For given
$\mu_1+\mu_2$ and $n_1,n_2$, the critical $q$ at which $e_{1,\rm eq}$
diverges is related to the characteristic eccentricity damping rate by
$q_{\rm crit}\propto t_{w,0}^{-2}$, where $t_{w,0}$ is a timescale
characterizing the migration and eccentricity damping defined as
$t_{\rm wave}$ [see Eq.~\eqref{twave}] evaluated at $m=m_1+m_2$ and
$a=a_2$.  Note that $t_{w,0}$ is determined by the disk parameters,
and is comparable to
$T_{e,0}$ of the larger planet.  This scaling for $q_{\rm crit}$ can be
explained as follows: 
The eccentricity of the smaller planet $e_{1,\rm eq}$ diverges when $\cos\theta_2\to 0$ according to \eqref{eq:varpi}.
When $\cos\theta_2\to 0$ (and $\sin\theta_2\to 1$), equation \eqref{eq:e2}, together with the fact that $e_{2,\rm eq}\sim
q^{1/2}h$, gives (assuming $q$ is small)
\eq{
\mu_1n_2 \sim e_{2,\rm eq}/T_{e,2}~~ \Rightarrow~~ q(\mu_1+\mu_2)n_2 \sim q^{1/2}h/t_{w,0},
}
which then gives $q_{\rm crit}\propto t_{w,0}^{-2}$.

A major caveat of the above calcuation is that the Hamiltonian \eqref{eq:Ham} and the
equations for the equilibrium, \eqref{eq:e1}-\eqref{eq:eta}, only
include the lowest-order terms in eccentricities; i.e., we have effectively
assumed $e_1,e_2\ll 1$. For realistic systems, when $e_1$ becomes
large, higher-order secular couplings may affect the result. We will
discuss this issue in the next subsection.

\subsection{Three-body simulations: effects of nonlinear eccentricities and non-adiabatic evolution}

\begin{figure}
\centering
\includegraphics[width=\linewidth]{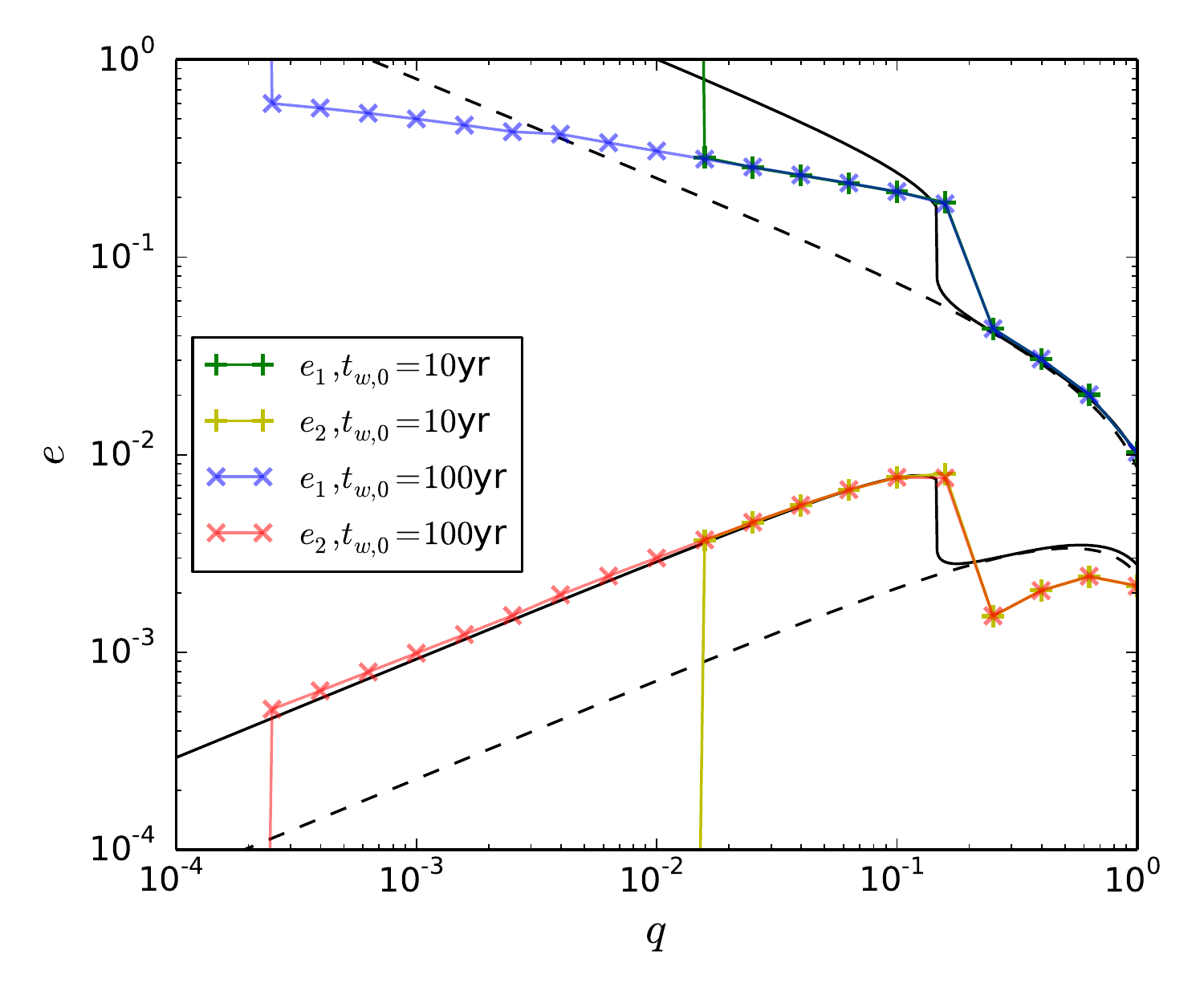}
\caption{Equilibrium eccentricities of two planets with
  $\mu_1+\mu_2=10^{-3}, M_\star=1M_\odot$ and initial $a_2=1$~au captured into the
  1:2 MMR in a disk with uniform surface density and $h=0.025$. The black
  curves show our analytical results, as given in Fig.~\ref{EqlibEcc_analytic}. 
  The 3-body integration results for $t_{w,0}=10$~yr (100yr)
  are shown in crosses (saltires). For $t_{w,0}=10$~yr, $e_1$ ($e_2$)
  is marked by the blue (red) curve; for $t_{w,0}=100$~yr, $e_1$ ($e_2$)
  is marked by the green (yellow) curve. When the inner planet is ejected
  (or collides with the other planet or the star), we set $e_1=\infty$
  and $e_2=0$.}
\label{numerical_12}
\end{figure}

\begin{figure}
\centering
\includegraphics[width=\linewidth]{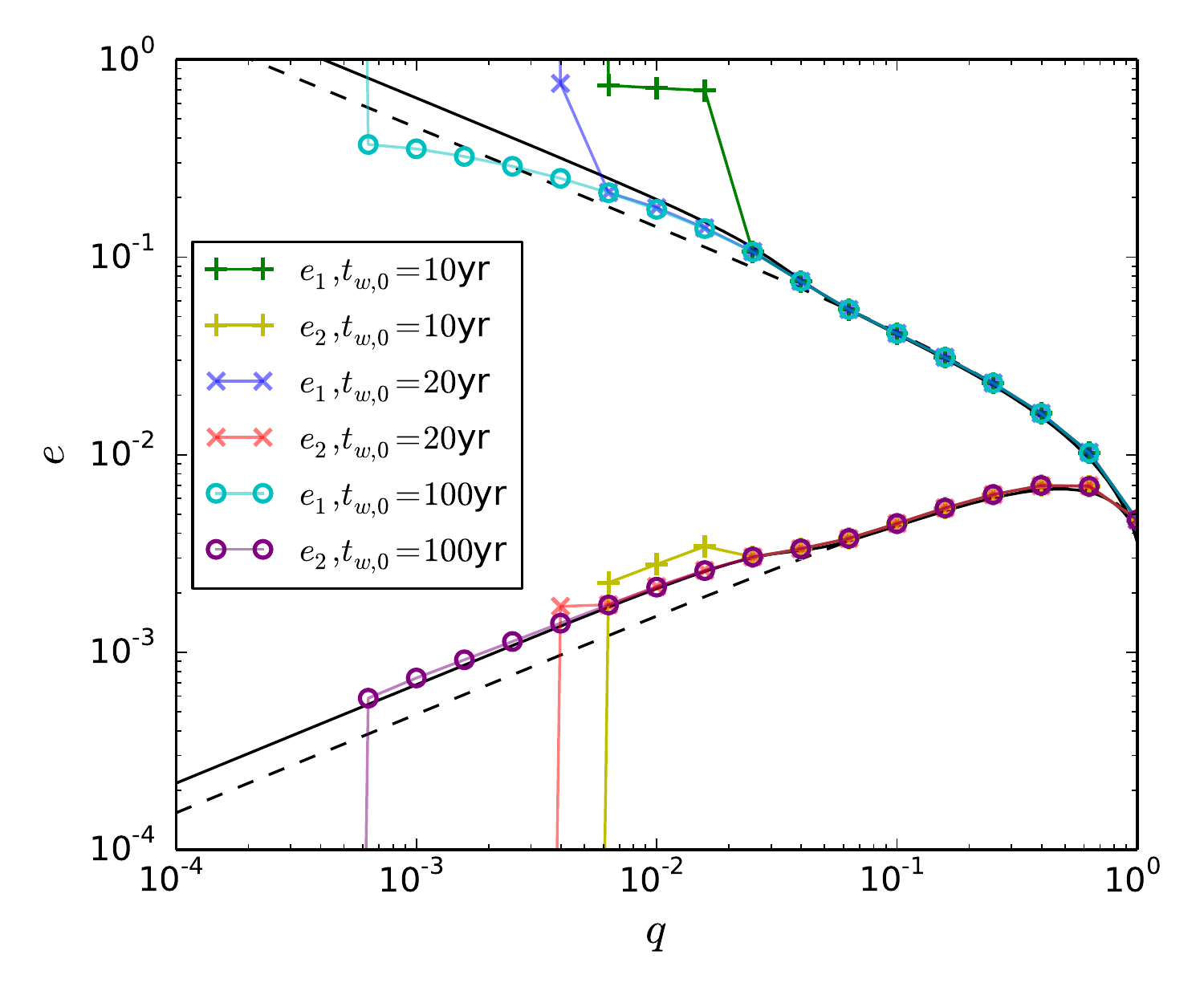}
\caption{Same as Fig.~\ref{numerical_12}, but for the 2:3 MMR. The
  behavior of the system is slightly different; see the text for more
  discussion.}
\label{numerical_23}
\end{figure}

\begin{figure}
\centering
\makebox[\linewidth][c]{\includegraphics[width=1.1\linewidth]{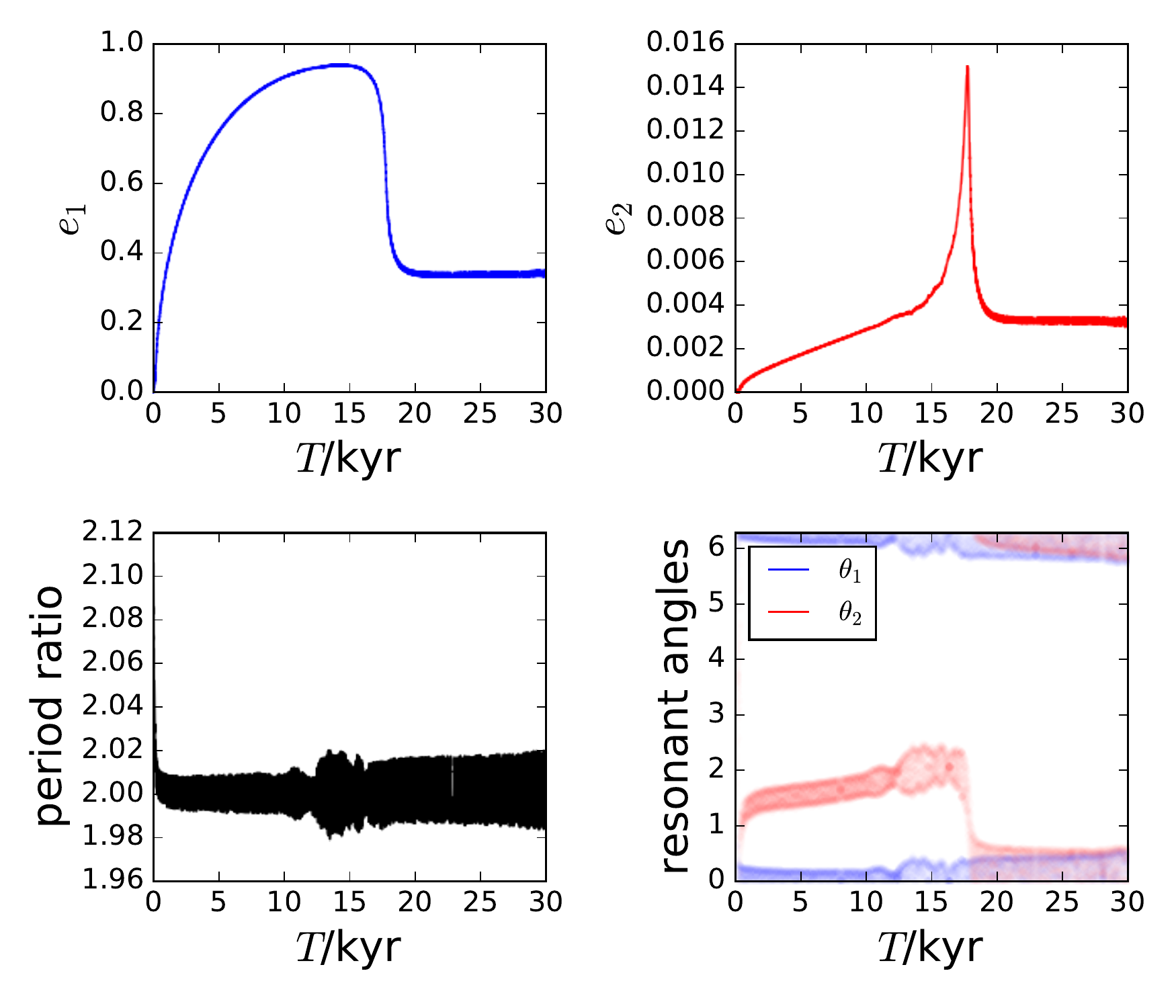}}
\caption{An example where the eccentricity $e_1$ overshoots to a large
  value before reaching equilibrium, for the system in
  Fig.~\ref{numerical_12} with $t_{w,0}=10$~yr and $q=0.125$. The different
  panels show the two planet's eccentricities ($e_1,e_2$), the period ratio
  and the resonant angles ($\theta_1=2\lambda_2-\lambda_1-\varpi_1$,
  $\theta_2=2\lambda_2-\lambda_1-\varpi_2$). In this example, $e_1$ reaches a maximum
  value of $0.935$ before decreasing to the equilibrium value. Note that for a
  slightly smaller $q$, the inner planet will have $e_1\to 1$ and
  become unstable during the overshoot.}
\label{overshoot}
\end{figure}

We now use 3-body simulations to check our semi-analytical results
obtained in the previous subsection.  This is necessary since the
Hamiltonian \eqref{eq:Ham} assumes that the eccentricities are small,
which may lead to nontrivial errors when $e_1$ attains large
values. In addition, it is useful to use 3-body integrations to
investigate at which point and for what reason(s) the inner planet
becomes dynamically unstable at high eccentricities.

Figures \ref{numerical_12} and \ref{numerical_23} compare the 3-body
integration results for the 1:2 and 2:3 MMRs using \textit{MERCURY}
with our analytical results. Forcing due to planet-disk interaction is
implemented as described in \citet{CresswellNelson08} to agree with
equations \eqref{Tm_act} and \eqref{Te_act}. Overall, the 3-body integration results
agree with our analytical results, showing the general trend
that the equilibrium eccentricities increase (compared to the simple
migration model) for small $q$.  However, there are several important effects
that the semi-anaytical linear theory fails to capture, and we discuss these
effects below.

\subsubsection{Effect of high-order coupling at large $e_1$}

Figure \ref{numerical_12} and the $t_{w,0}=100$yr curves\footnote{The other curves in Figure
\ref{numerical_23} will be discussed in Section \ref{subsubsec:eq_bifurcation}.}
in Figure \ref{numerical_23} show that $e_{1,\rm eq}$ is
smaller than the analytical prediction when $e_{1,\rm eq}\sim 1$.
This is likely due to the higher-order secular coupling between the
planets; such coupling prevents $e_1$ from reaching unity while $e_2$
remains finite.  As a result, the divergence of $e_1$ due to finite
eccentricity damping (discussed in Section
\ref{subsubsec:finite_damping}) does not occur in real systems. (The
ejection of the inner planet for small $q$ depicted in Figure
\ref{numerical_12} and Figure \ref{numerical_23} are due to eccentricity
overshoot, a phenomenon we will discuss next.)

\subsubsection{Effect of non-adiabatic evolution: eccentricity overshoot}

For sufficiently slow migration, 
the evolution of the system is adiabatic 
(i.e. the evolution of $\eta$, the ``resonance depth" parameter,
is sufficiently slow so that the system stays close to the libration center as
the libration center moves in the phase space)
and the eccentricities of both planets should slowly
increase until they reach the equilibrium values.  In this case, the
equilibrium eccentricities are the maximum eccentricities that the planets
can reach.  However, when $q$ is small or when migration is fast
(i.e. $t_{w,0}$ is small), the growth of $e_2$ is too slow, and the initial evolution
of $e_1$ is similar to the restricted problem studied in Section 3:
Due to the inefficient eccentricity damping, $\eta$ and $e_1$ both keep increasing,
and $e_1$ can easily grow beyond the equilibrium value.
The growth of $e_1$ stops only when it becomes so large that the secular interaction
between the planets forces $e_2$ to increase. Since eccentricity damping of $e_2$
is still efficient, this stops the system from going deeper into the resonance (i.e. stops $\eta$ from further increasing).
Eventually, the system will reach equilibrium, provided that the smaller planet
has not become dynamically unstable during
the high-$e_1$ phase.

Figure \ref{overshoot} shows an
example. Before the system reaches equilibrium, the eccentricity $e_1$
first overshoots to a very large value, then decreases back to the
equilibrium value.  When $q$ is smaller (or when the migration is
faster), the inner planet will be ejected because it reaches $e_1\to
1$ during this overshooting phase. This is the reason for the ejection
of the smaller planet at low $q$ in Figures \ref{numerical_12} and
\ref{numerical_23}.

It is worth noting that significant eccentricity overshoot is a
phenomenon unique to the realistic migration model. For the simple
migration model, since the eccentricity damping of the inner planet is
efficient (i.e. $e_1^2/T_{e,1}$ always increases as $e_1$ increases),
the system will cease to go deeper into the resonance once the $e_1^2$
terms in equation \eqref{eq:eta} can balance the migration;
this corresponds to an insignificant eccentricity overshoot.

\subsubsection{Effect of non-adiabatic evolution: bifurcation of the equilibrium state}\label{subsubsec:eq_bifurcation}

In Figure \ref{numerical_23}, we observe that 
the equilibrium eccentricity of the small planet increases abruptly 
when $q$ goes below $q\simeq 0.02$ (0.005) for $t_{w,0}=10$~yr
(20~yr); at a somewhat smaller $q$ the system becomes unstable. It is
likely that this abrupt change corresponds to a bifurcation, with the
equilibrium states before and after the bifurcation corresponding to
two different fixed points of the system.  One possible reason for
this bifurcation is that the finite migration rate, together with the
more realistic migration model, affect the stability of the fixed
points.  This different equilibrium state with a higher equilibrium
eccentricity is not captured by our analytical result. Also, for this
new equilibrium state we observe less eccentricity overshoot.

As $t_{w,0}$ increases, the intermediate region where the system
reaches this different equilibrium state with high
eccentricity shrinks; when $t_{w,0}$ is sufficiently large the
system always becomes unstable (due to eccentricity overshoot) before
the bifurcation happens and this intermediate region disappears.

\subsection{Stability of capture}

\begin{figure}
\centering
\makebox[\linewidth][c]{\includegraphics[width=1.1\linewidth]{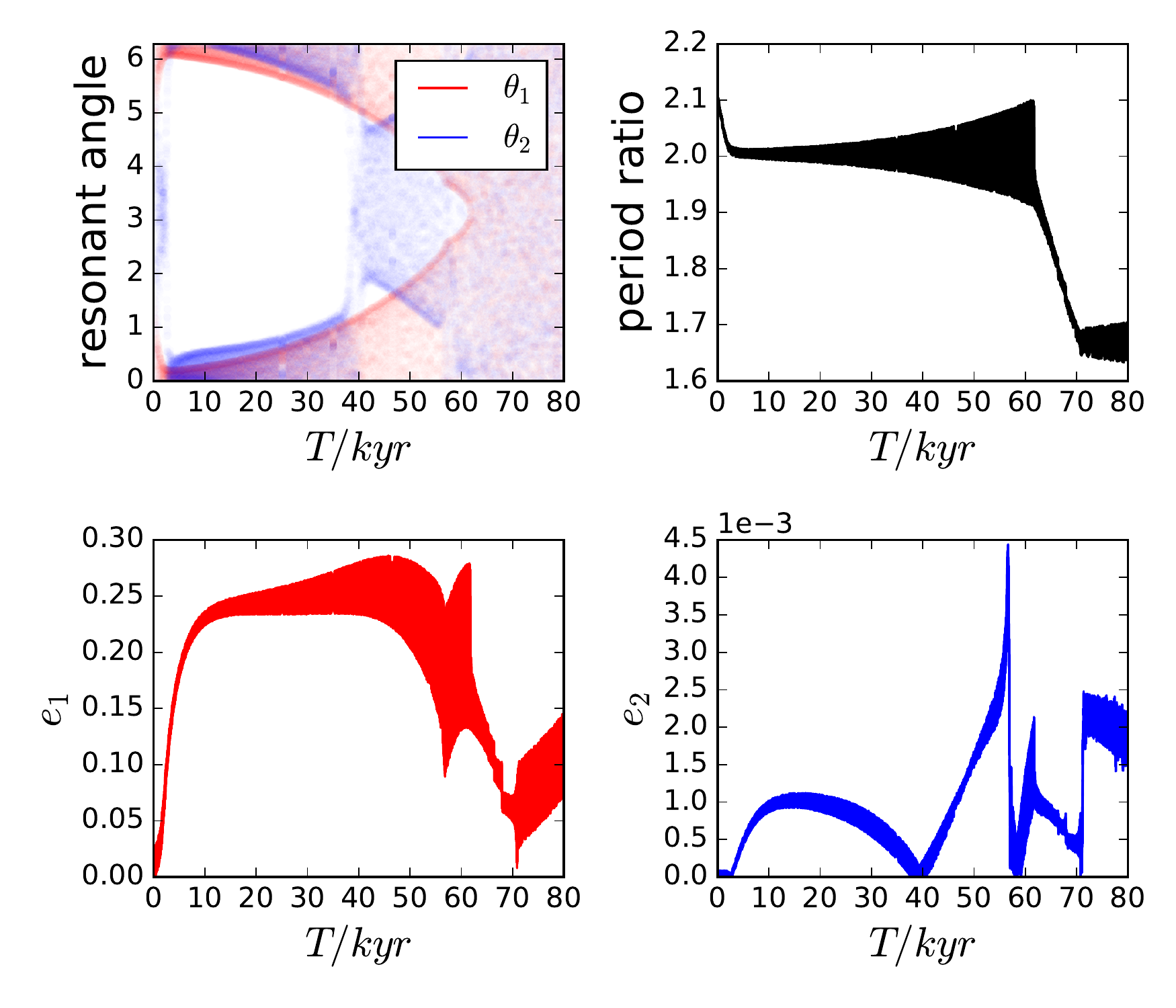}}
\caption{Outcome of the 1:2 MMR capture for the system depicted in
  Fig.~\ref{numerical_12} with $q=0.01$ and $t_{w,0}=100$~yr, using the
  simple migration model. We see that the system escapes the resonance at
  $t\simeq 60$~kyr.}
\label{massive_simple}
\end{figure}

\begin{figure}
\centering
\makebox[\linewidth][c]{\includegraphics[width=1.1\linewidth]{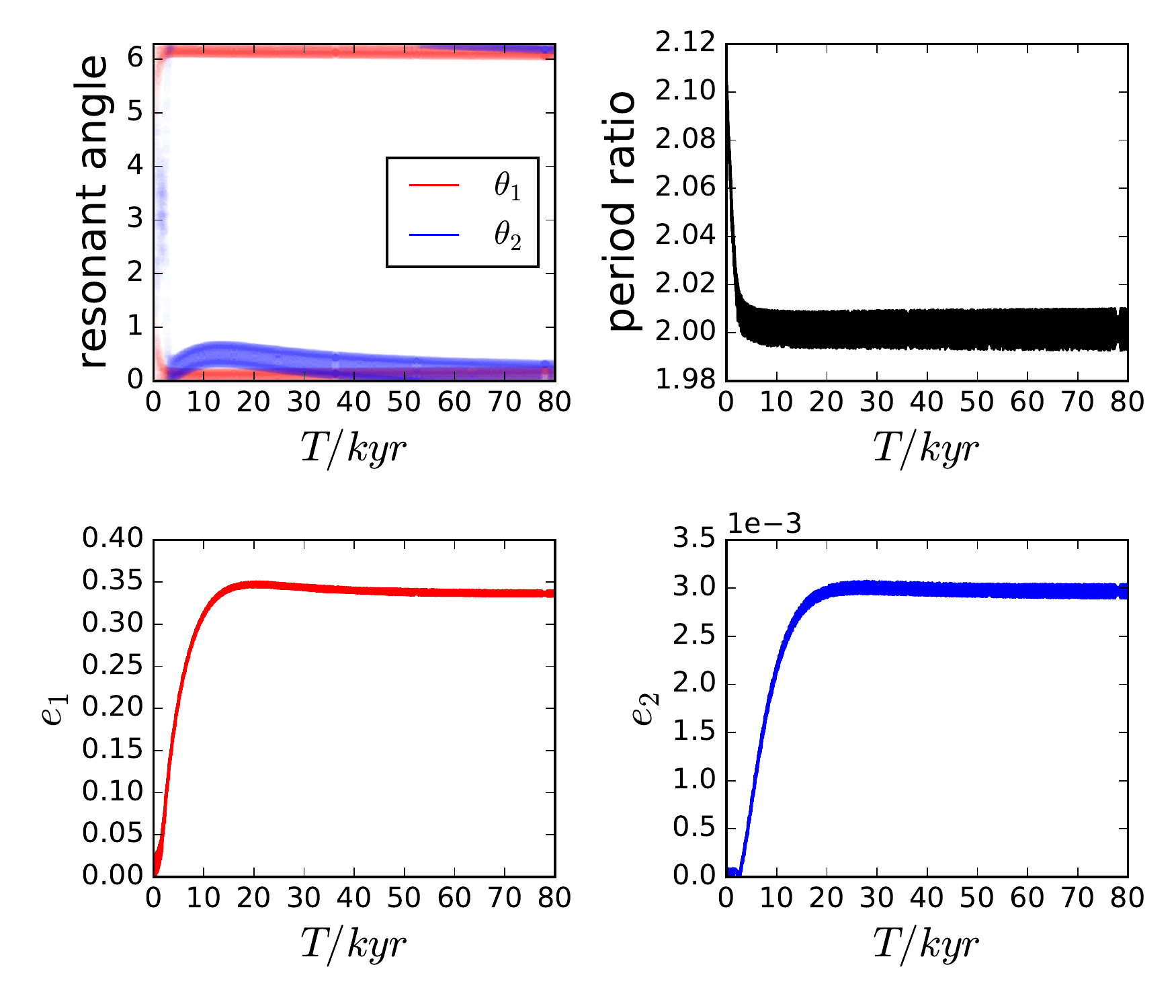}}
\caption{Same as Fig.~\ref{massive_simple}, except that the realistic
  migration model is used. The equilibrium state is stable.  Note that
  the equilibrium eccentricity also is increased compared to
  Figure \ref{massive_simple}.  }
\label{massive_realistic}
\end{figure}

Similar to the case when the smaller planet is massless (Section 3),
using the realistic migration model affects the stability of MMR
capture. We observe that when the equilibrium eccentricity is
$\gtrsim$ a few $h$, the system tends to be more stable compared to
the prediction of the simple migration model. Since it is
difficult to do a thorough survey of the parameter space, we
illustrate this by an example. Figures \ref{massive_simple} and
\ref{massive_realistic} show the different outcomes of a 1:2 MMR
capture when the simple migration model (eccentricity-independent
$T_e$ and $T_m$) and the realistic migration model
(eccentricity-dependent $T_e$ and $T_m$) are used.  For the simple
migration model, the equilibrium state is overstable, and the system
eventually escapes the resonance.  For the realistic migration model,
the equilibrium state becomes stable (the eccentricity at the
equilibrium also increases compared to the simple migration model).

Moreover, all numerical examples summarized in
Figs.~\ref{numerical_12} and \ref{numerical_23} (except 
those caese where the inner planet is ejected) have stable equilibrium
states. This suggests that for planets undergoing type-I migration,
the capture into a first-order MMR is stable for typical
disk configurations if we use the realistic migration model. By
contrast, if we use the simple migration model for the systems in
Figs.~\ref{numerical_12} and \ref{numerical_23}, the equilibrium state
becomes unstable for $q\lesssim 0.1$.

\citet{DeckBatygin15} have previously carried out an extensive study
on the stability of the equilibrium state of first-order MMRs for
general planet mass ratios. Their analysis was entirely based on the
simple migration model.
They found a region of the parameter space leading to overstability and proposed a 
criterion for overstability of the equilibrium state. 
Since an overstable system tends to evolve to an adjacent MMR equilibrium state,
they concluded that the overstability of the
equilibrium state cannot fully explain the observed paucity of
resonant pairs in the Kepler sample.

However, the overstability criterion of \citet{DeckBatygin15} cannot
be directly generalized to the realistic migration model (with
eccentricity-dependent $T_m,T_e$). This is because the stability of
the equilibrium state depends on both $\dot a /a, \dot e/e$ at the
equilibrium and their partial derivatives with respect to the
eccentricity.  Although it is possible to tune the parameters of the
simple migration model ($T_{e,0},~T_{m,0}$ for each planet and $p$) to
obtain $\dot a /a, \dot e/e$ and $\partial(\dot a /a)/\partial e$ that
locally match the values for the realistic migration model near the equilibrium,
the local value of $\partial(\dot e /e)/\partial e$ in general cannot
be matched by tuning the parameters of the simple migration model.
Still, if we oversimplify the problem by plugging the local values of
$T_e,~T_m$ at the equilibrium into the overstability criterion of
\citet{DeckBatygin15}, the stability does tend to increase compared to
the simple migration model (with $T_e=T_{e,0},~T_m=T_{m,0}$) when
$e_1\gtrsim$ a few $h$. This is mainly because $T_{e}$ of the inner
planet for the realistic migration model is larger than that for the
simple migration model, which pushes the system away from the
instability zone (see Figures 2 and 3 of \citealt{DeckBatygin15}). Note
that this is only an intuitive explanation of our finding of
the increased stability and cannot serve as a rigorous analysis.


\section{Summary and discussion}

\subsection{Summary of key results}
In this paper we have carried out theoretical and numerical studies on
the outcomes of first-order MMR capture for planets undergoing convergent type-I
migration. Unlike previous works \citep{GoldreichSchlitchting14,
  DeckBatygin15, Delisle15, XuLai17} which adopted a simple migration
model where the eccentricity damping rate and orbit decay rate
[$T_e^{-1}$ and $T_m^{-1}$ respectively, see equations \eqref{eq:lin1}
  and \eqref{eq:lin2}] are independent of the planet's eccentricity,
we consider a more realistic model for $T_e$ and $T_m$ which captures
their nonlinear eccentricity dependence when the eccentricity exceeds $\sim h$
(where $h\equiv H/r$ is the aspect ratio of the disk).
We find that this more realistic migration model can significantly
affect the outcomes of MMR capture and lead to several new dynamical
behaviors.

First, the equilibrium eccentricities of planets captured into the MMR
can be larger by a factor of a few than those predicted by the simple
migration model (which assumes eccentricity-independent $T_e,T_m$).
This arises because when $e\gtrsim 3h$, eccentricity damping becomes
weaker and the system migrates deeper into the resonance before
reaching equilibrium.  When the inner planet is massless, the
equilibrium state no longer exists if the equilibrium eccentricity
predicted using the simple migration model is $\gtrsim 3h$, and the
planet's eccentricity grows and eventually becomes unstable (Section
3.1). For general planet mass ratios (section 4), the more massive planet's eccentricity
stays below $3h$, and the eccentricity damping of this more massive
planet ensures the existence of the equilibrium state.  However, the
equilibrium eccentricity is larger than the prediction using the
simple migration model when the mass ratio $q=m_1/m_2$ is sufficiently
small (Section 4.1). This increase in eccentricity is very significant for the 1:2
MMR, and less significant for other first-order MMRs (see Fig.~8).
For typical disk parameters, the critical mass ratio below which such
increase occur is around $0.03-0.15$ (see Figs.~8-9).

Second, the stability of the equilibrium state can be strongly
affected by the migration model.  Our analytical calculation and
parameter survey for the case when the inner planet is massless (Section 3) show
that the equilibrium state becomes more stable when the equilibrium
eccentricity is $e_{\rm eq}\gtrsim 2h$ (Section 3.2; see Fig.~3). This
increased level of stability of MMR is also seen when both planets
have finite masses (Section 4.3).  In particular, for realistic disk
configurations, the simple migration model predicts that the
equilibrium state is unstable for small $q$, while the realistic
migration model predicts that the equilibrium state is virtually
always stable 
(provided that the small planet does not suffer dynamical ejection at 
high eccentricities; see below).

Another new phenomenon we have found is that when the migration is fast
and/or the inner planet's mass is sufficiently small, the eccentricity growth of the more
massive planet (due to the resonant perturbation from the inner, smaller 
planet) becomes too slow; this causes the eccentricity of the smaller
planet to overshoot the equilibrium value before the system reaches
the equilibrium state (Section 4.2.2; see Fig.~13). Such an overshoot
can be very significant and may cause the smaller planet to be ejected
at high eccentricities even when the equilibrium eccentricity is modest.

Overall, using the more realistic migration model tends to increase
the equilibrium eccentricities of planets captured in MMRs and make
the equilibrium state less prone to overstability. However, when migration is 
sufficiently fast 
(or the small planet has too small a mass), 
it also causes the ejection of the smaller planet during
eccentricity overshoot --- this behavior is much less significant when the
simple migration model is used. All of these can affect the ways in
which MMRs shape planetary system architecture.

\subsection{Implications for multi-planet system architecture}

\subsubsection{Occurrence of MMRs} 

For planets with similar masses ($q\sim 1$), since the equilibrium
eccentricities of the planets captured into MMRs are usually small,
previous results concerning the stability of MMRs remain valid 
\citep{DeckBatygin15,Delisle15,XuLai17}.

For smaller mass ratio ($q\lesssim 0.1$), however, the maximum
eccentricity that the smaller planet can reach is much larger when the
realistic migration model (with eccentricity-dependent $T_e,~T_m$) is
applied (compared to the results obtained with
eccentricity-independent $T_e,~T_m$) due to the increased equilibrium
eccentricity and eccentricity overshoot.  The large eccentricity can
lead to the ejection of the smaller planet (when its eccentricity approaches
unity) or make it scatter with a third planet in the system (if their
orbits cross).  This tends to reduce the multiplicity of the system
when it initially hosts a pair of convergently migrating planets with
small mass ratio. This effect also reduces the number of small-mass-ratio
planet pairs in MMRs.

\subsubsection{Loneliness of Hot Jupiters} 

The eccentricity overshoot phenomenon (which occurs when the
mass ratio is small and the migration is sufficiently fast) provides
an efficient way of removing super-Earth companions of fast migrating
giant planets. This may help explain the loneliness (the lack of
low-mass planet neighbors) of hot Jupiters \citep{Huang16}
if they are formed through disk-driven migration.\footnote{
Giant planets should undergo type-II instead of
  type-I migration. Although in this paper we have focused on type-I
  migration models, our results should also be reasonably accurate if
  the more massive planet undergoes type-II migration, since the
  eccentricity dependence of the more massive planet's migration and
  eccentricity damping rates does not play an important role in our
  analysis.}
In this picture, hot Jupiters arrived at their current
locations through fast type-II migration, with the migration timescale
much less than the disk lifetime\footnote{Type II migration can be
  fast enough to push a Jupiter to the disk inner edge before the gas
  disperses \citep{HasegawaIda13}. Some of these hot Jupiters may not fall into the star
  probably because their migration stop when reaching the inner edge
  of the disk or when the inner part of the disk induces outward
  migration (e.g. \citealt{Lega15}).}. If they had any inner low-mass companion
(a super-Earth), it could be removed when captured into a
MMR with the Jupiter during its migration due to the instability
caused by eccentricity overshoot. On the other hand, warm Jupiters 
do commonly have low-mass planet companions (Huang et al.~2016).
This may be explained by their slow migration rates:
During such slow migration, their low-mass companions do not suffer 
eccentricity overshoot and therefore are kept in safety upon capture into 
MMRs. Note that the rate of type-II migration is sensitive to the property
of the disk, especially its viscosity \citep{Ward97}. Thus, in this scenario,
whether a system forms hot Jupiters (without low-mass companions) or warm Jupiters
(with low-mass companions) simply reflects the different disk properties and 
the resulting different migration history of giant planets.

Of course, hot Jupiters may also form by high-eccentricity migration, in which
the eccentricity of a giant planet is excited by distant stellar or planetary companions,
followed by tidal circularization and orbital decay (e.g. \citealt{DawsonJohnson18}).
In this scenario, the loneliness of hot Jupiters can be naturally
explained because a giant planet undergoing high-amplitude
eccentricity oscillations can easily eject smaller planets interior of
its initial orbit.\footnote{However, in this scenario, warm Jupiters with
inner companion should be formed through a different channel.}  Our
discussion here does not aim to prove or disprove any particular
formation scenario; we simply argue that one should not rule out the
disk-driven (low-eccentricity) migration scenario using solely the
loneliness of Hot Jupiters.

\section*{Acknowledgments}
We thank M. Lambrechts, E. Lega and G. Pichierri for useful discussions and comments.
This work has been supported in part by NASA grant NNX14AG94G 
and NSF grant AST-1715246. WX thanks the undergraduate research fellowship from the 
Hopkins Foundation (Summer 2017). DL thanks the Laboratoire Lagrange OCA for 
hospitality where this work started.

\bibliographystyle{mnras}
\bibliography{BIB}
\end{document}